\documentclass[twocolumn,smallcondensed]{svjour3}     

\smartqed 
\usepackage{graphicx}
\usepackage{color}
\usepackage{booktabs}
\usepackage[table,xcdraw]{xcolor}
\usepackage{algorithm,algpseudocode}
\usepackage{float}
\usepackage{amssymb}
\usepackage{amsmath}
\usepackage{natbib}
\usepackage{lineno}
\usepackage{soul}
\usepackage{todonotes}
\usepackage{bm}
\usepackage{subfig}
\usepackage{siunitx}
\usepackage[misc]{ifsym}

\usepackage{breakcites}
\usepackage[colorlinks=true,
            linkcolor=red,
            urlcolor=blue,
            citecolor=blue,
            breaklinks=true]{hyperref}%

\usepackage{soul}
\usepackage{comment}

\begin{document}



\title{Analysis of an unsteady quasi-capillary channel flow with Time Resolved PIV and RBF-based super resolution}

\author{Manuel Ratz, Domenico Fiorini, Alessia Simonini, Christian Cierpka and *Miguel A. Mendez}
\authorrunning{Ratz M. \and Fiorini D. \and Simonini A. \and Cierpka C.\and Mendez M.A.}


\institute{M. Ratz, C. Cierpka \at Technische Universität Ilmenau, Institute of Thermodynamics and Fluid Mechanics, Germany\\
              \and
D. Fiorini, A. Simonini, M.A. Mendez (\Letter) \at
von Karman Institute for Fluid Dynamics, Belgium
\\
 \email{mendez@vki.ac.be}
}


\date{\today}

\maketitle

\begin{abstract}
We investigate the interface dynamics in an unsteady quasi-capillary channel flow. The configuration consists of a liquid column that moves along a vertical 2D channel, open to the atmosphere and driven by a controlled pressure head. Both advancing and receding contact lines were analyzed to test the validity of classic models for dynamic wetting and to study the flow field near the interface. The operating conditions are characterized by a large acceleration, thus dominated by inertia. The shape of the moving meniscus was retrieved using Laser-Induced Fluorescence (LIF)-based image processing while the flow field near was analyzed via Time-Resolved Particle Image Velocimetry (TR-PIV).
The TR-PIV measurements were enhanced in the post-processing, using a combination of Proper Orthogonal Decomposition (POD) and Radial Basis Functions (RBF) to achieve super-resolution of the velocity field. Large counter-rotating vortices were observed, and their evolution was monitored in terms of the maximum intensity of the $Q$-field. 
\\The results show that classic contact angle models based on interface velocity cannot describe the evolution of the contact angle at a macroscopic scale. Moreover, the impact of the interface dynamics on the flow field is considerable and extends to several capillary lengths below the interface.



\end{abstract}

\section{Introduction} \label{sec:introduction}

The dynamics of a gas-liquid interface moving along a wall play an essential role in many industrial applications such as wetting and dewetting processes (e.g. inkjet printing, see \cite{eral2013contact}), coating applications (e.g. slot die coating, see \cite{kistler1997liquid}) or capillary driven flows (e.g. liquid absorption in porous media, see \cite{ma2012visualization}). 

\begin{figure*}[htp]
    \centering
    \subfloat[]{\label{fig:split_injection}
      \includegraphics[height=0.17\paperheight]{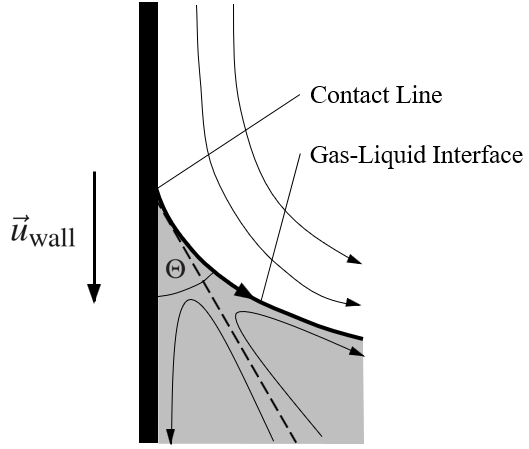}}
~
    \subfloat[]{\label{fig:split_ejection}
      \includegraphics[height=0.17\paperheight]{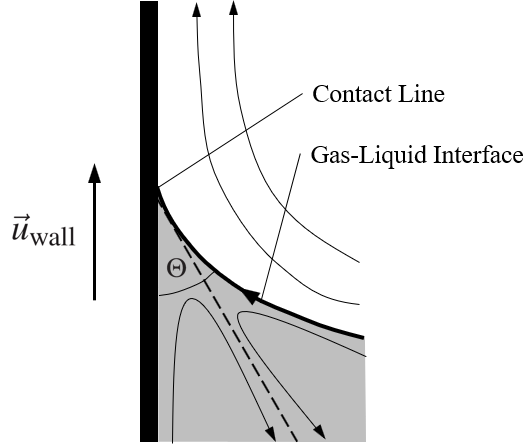}}

    \caption{Schematic sketch of the stream lines in proximity of the contact line for a wetting configuration ($\theta<90^o$) in case of an advancing contact line (left) and a receding contact line (right). The flow pattern on the left is known as \emph{stream line split injection}, the flow pattern on the right is known as \emph{stream line split ejection}. Images adapted from \cite{fuentes2005}.}\label{fig:split_stream}
\end{figure*}

Sufficiently close to the contact line, i.e. the intersection of the gas-liquid interface and the solid bounding walls, capillary forces play a leading role in the dynamics of the interface motion and thus the liquid spreading. At a macroscopic level, capillary forces depend on the curvature of the gas-liquid interface, but their computation requires boundary conditions at the wall: these are usually given in terms of \emph{contact angle} $\Theta$, that is the angle between the gas-liquid interface and the solid wall at the contact line (cf. Figure \ref{fig:split_stream}). 

Because the contact angle cannot be predicted in the framework of continuous mechanics, one must resort to non-equilibrium statistical mechanics, \emph{ad hoc} assumptions in the continuous framework \citep{Dussan1976,COX,Voinov1977} or empirical laws that link the contact angle to macroscopic quantities such as the contact line velocity (e.g. \cite{Hoffman75,Kistler93}) or acceleration (e.g. \cite{bian2003liquid,Ting1995}). \textcolor{black}{While these approaches have been successful in `quasi-steady' conditions and very viscous fluids \citep{quere1997inertial,Wu2017}, their generalization to accelerating contact lines and low-viscosity fluids remain an open research topic for both capillary or quasi-capillary flows (see \cite{fiorini2021,quere1997inertial, shardt2014inertial,willmott2020inertial}) and impacting droplets (\cite{BARTOLO2005,Wang2020})}.

A well known challenge in the modelling of flows near contact lines is the singularity that arises at the contact line when combining the notion of \emph{no-slip}, i.e. zero velocity at the wall, with the notion of \emph{dynamic contact line} \citep{dussan1971,huh1971}. A possible resolution to this paradox is the splitting streamline pattern proposed by \cite{huh1971}. This pattern can be modeled as a \emph{creeping flow} obeying a bi-harmonic equation in the stream-function and is schematically illustrated in Figure \ref{fig:split_stream} for advancing and receding contact lines. In an advancing configuration (Figure \ref{fig:split_injection}), the fluid is pushed from the bulk of the flow towards the wall, while in a receding configuration (Figure \ref{fig:split_ejection}), the opposite pattern is expected. In both cases, the splitting streamlines pattern produces a rolling motion near the contact line and has been observed in various experimental campaigns (\cite{chen1997,fuentes2005,nasarek2008,zimmermann2010}). 
In an open channel flow, such as those encountered in capillary-rise problems \citep{washburn1921,levine1976}, the presence of the rolling motion collides with the assumption of fully developed flows. It is thus relevant to analyze how far from the interface one might expect to see the impact of capillary forces near the contact line \citep{savelski1995,nasarek2008}.

\begin{figure*}[htp]
    \centering
    \subfloat[]{\label{fig:exp_theory}
      \includegraphics[width=.43\textwidth]{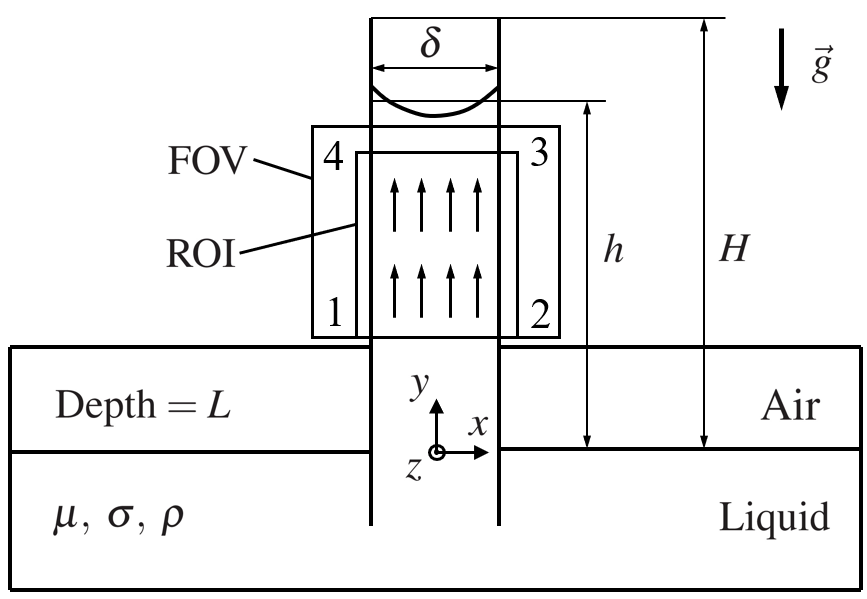}}
~
    \subfloat[]{\label{fig:exp_piv}
      \includegraphics[width=.43\textwidth]{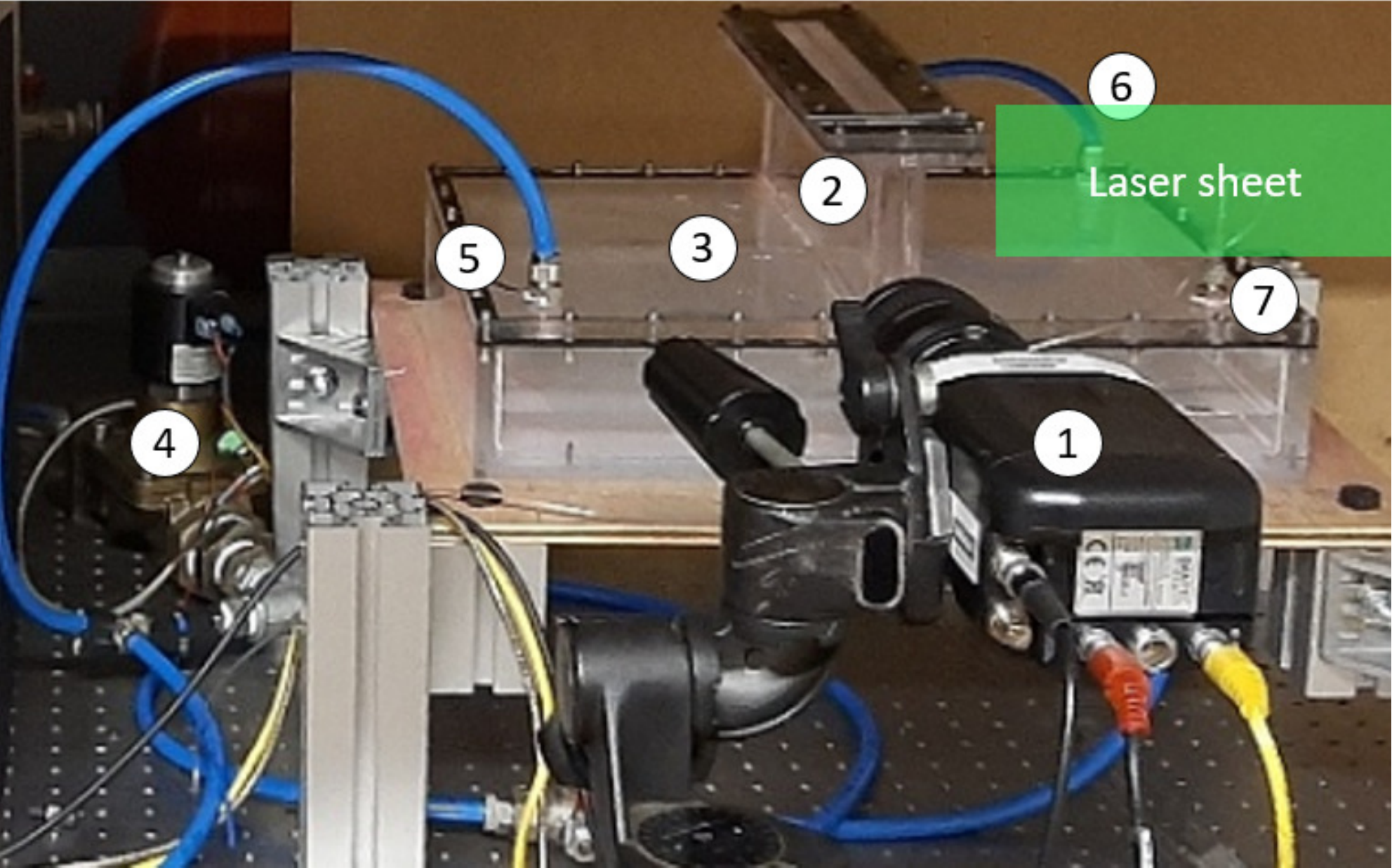}}

    \caption{Fig (a) Sketch of the channel showing the area of the TR-PIV measurements. The Field Of View (FOV) and Region of Interest (ROI) are discussed in detail in Section \ref{sec:piv_setup}. Fig (b) shows a photography of the experimental set-up with the camera (1), the channel (2), the reservoir (3), the release valve (4), the pressure ports for the injection (5 and 6) and discharge (7).}\label{fig:experimental_setup}
\end{figure*}

The scope of this work is twofold. On the one hand, we aim at analyzing the well-posedness of a correlation linking the macroscopic contact angle to the kinematics of the contact line in inertia dominated conditions, i.e. in the presence of large accelerations and large velocities. On the other hand, we aim to analyze the flow field in the proximity of the contact line and measure up to which distance from the interface the rolling motion is present. Both aspects are analyzed considering a 2D, quasi-capillary channel (i.e. with width $\delta\approx 2l_c$, with $l_c=\sqrt{\sigma/\rho g}$ the capillary length, $\sigma$ the surface tension, $\rho$ the liquid density and $g$ the gravitational acceleration). The experimental set-up and measurement conditions are recalled in Section \ref{sec:experimental_setup}. The configuration of interest is a narrow rectangular channel in which water moves along a solid wall forming two dynamic contact angles. The flow is sustained by a controlled pressure head that reproduces both advancing and receding contact lines.

The shape of the interface was tracked using Laser-Induced Fluorescence (LIF)-based image visualization (\cite{ledar_original}) and edge detection, while the contact line was characterized via an inverse method by fitting a simple model for the interface shape (\cite{fiorini2021}). The set-up for the LIF-based flow visualization and the related image processing is described in Section \ref{sec:lif_setup}. The velocity field was characterized via Time-Resolved Particle Image Velocimetry (TR-PIV). Vortices near the contact line were visualized and their strength quantified in terms of $Q$-fields.

To accurately compute the spatial derivatives required in the $Q$-field definition, we propose a super-resolution method that combines Proper Orthogonal Decomposition (POD) and Radial Basis Function (RBF) regression (\cite{kkarri2009,raben2012}). The set-up for the TR-PIV measurements and the post processing of the velocity fields is described in Section \ref{sec:piv_setup}. The results of both experimental campaigns are presented in Section \ref{sec:results} while conclusions and perspectives are collected in Section \ref{sec:conclusion}.

\section{Experimental Set-up and Test Cases}
\label{sec:experimental_setup}

 A sketch of the experimental set-up for the measurements is shown in Figure \ref{fig:exp_theory}. This consists of a rectangular channel with width $\delta = \SI{5}{\milli\meter}$, in the figure depth $L = \SI{250}{\milli\meter}$ and height $H = \SI{150}{\milli\meter}$.

This channel is open to atmosphere at the top and connected to a pressurized chamber on the bottom. The chamber sustains the flow via \textcolor{black}{a (time varying) gauge pressure $\Delta p(t)$. To reproduce an advancing contact line (i.e. interface moving upwards, along the $y$-axis), a pressure step of amplitude $\Delta P$ is introduced by the opening of a fast electronic valve connected to a pressure line. To reproduce a receding contact line (i.e. interface moving downward), the same procedure is followed in reverse: the chamber is initially pressurized, i.e. $\Delta p(0)> 0$ and the release valve is opened to the atmosphere.}

\begin{figure}
     \centering
     \includegraphics[width=.47\textwidth]{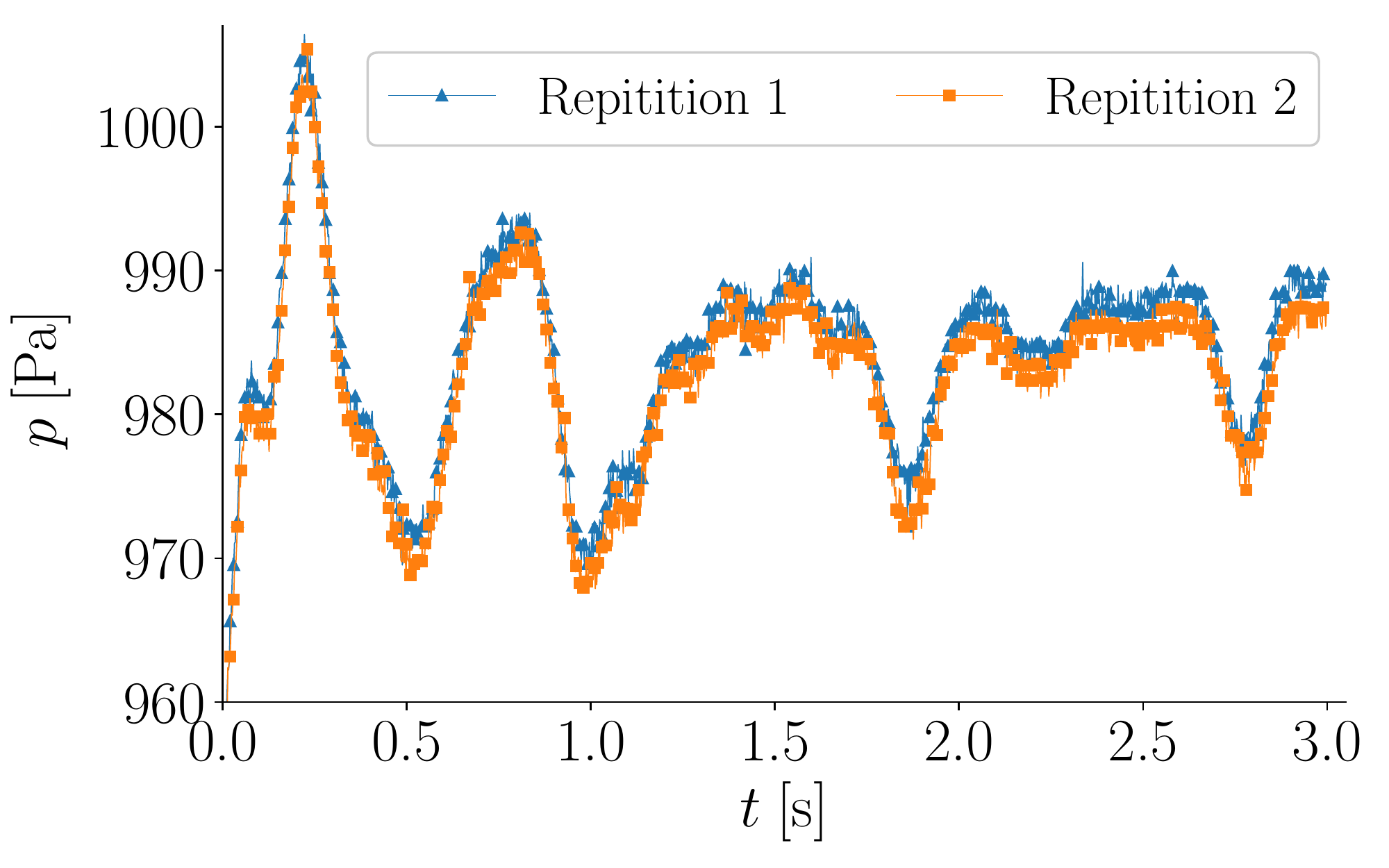}
     \caption{Temporal evolution of the gauge pressure in the pressure tank with a pressure difference $\Delta$P = \SI{1500}{\pascal} for two different experiments.}\label{fig:pressure_plots}
 \end{figure}

In the current set-up, the opening of the valve is carried out manually. Therefore, the experimental conditions for the receding contact line can slightly differ between experiments. Nevertheless, the time evolution of the pressure inside the tank is monitored with a pressure gauge and can thus be linked to the dynamics of the interface motion as described in Section \ref{sec:Model}. For the advancing contact angle problem, the fast opening of the valve (which occurs linearly in about $0.2$ s) makes the evolution of the pressure in the channel more repeatable, as shown in Figure  \ref{fig:pressure_plots}. 

A picture of the experimental setup is shown in Figure \ref{fig:exp_piv}. The high speed camera is placed perpendicularly to the channel’s cross section, which is illuminated with a laser sheet from the right side. The camera and the laser sources changed between the LIF-visualization and the PIV experiments, as described in Sections \ref{sec:lif_setup} and \ref{sec:piv_setup} respectively.

 The injection/removal of air during the pressurization/depressurization used in the advancing/receding experiments was performed on two sides of the chamber to ensure the symmetry of the flow and minimize the entry effects. Together with the camera (1), the picture shows the channel (2) and the pressurized chamber (3), the fast opening valve (4), the two pressure ports (5 and 6) and the pressure gauge (7). 
 
 \textcolor{black}{The walls of the channel were made in acrylic glass and the experiments were carried out using demineralized water (density $\rho=\SI{997}{\kilo\gram\per\meter^3}$, dynamic viscosity $\mu=\SI{0.001}{\pascal\second}$, and surface tension $\sigma=\SI{0.072}{\milli\newton\per\meter}$). The liquid was introduced into the facility via a lateral port, and all experiments were carried out on an initially dry surface. The influence of the surface pre-conditioning was tested by repeating the advancing experiments after intervals of approximately $15$ minutes. Since no noticable difference was observed in the interface dynamics (nor in the contact evolution) between the first and the following experiments, we conclude that this time is long enough to have dry surfaces. }

 \textcolor{black}{The static contact angle was measured using the same LIF-based image processing approach described in Section \ref{sec:lif_setup} on a quiescent meniscus. This approach is analogous to the Meniscus Profile Method (MPM) as proposed by \cite{petrov1993quasi}. The interface was positioned in the camera's field of view by gently rising the chamber pressure, and the measurement was taken after waiting long enough to have a static interface. The static contact angle was found to be $\Theta_0=33 ^o\pm2 ^o$.} 
 
Finally, we emphasize that the need for splitting the experimental campaigns into a high-speed LIF and a TR-PIV stems from the different objectives and the different technical constraints for the two measurements. The LIF-based visualization aims to analyze the motion of the gas-liquid interface during its rise and oscillation, while the TR-PIV seeks to explore the flow field during the passage of the interface.
Consequently, the LIF-based visualization required videos of longer duration, lower frame rate and larger field of view than the TR-PIV campaign. Before proceeding with the description of the measurement set-up, it is worth elaborating on a simple attempt to model the response of the liquid column in the next section; this also allows to better understand the test conditions investigated.

 \subsection{Modeling and Parameter Definition}\label{sec:Model}

 Models of the capillary rise have been proposed by \cite{levine1976,washburn1921} and were used for the current campaign to have a preliminary evaluation of the interface displacement as a function of the imposed pressure head. 

Denoting as $m$ the mass of the water column in motion and as $h(t)$ the liquid height as a function of time, the force balance acting at the inlet of the channel sets 

 \begin{align}
     \label{eq:momentum}
     \frac{\text{d}}{\text{dt}}\left(m \dot{h}\right) = F_g + F_{\Delta p} + F_\text{s} + F_\text{v}\,
 \end{align} where $\dot{h}$ is the time derivative of the column height (thus the mean velocity of the flow), $F_g$ is the force due to gravity, $F_{\Delta p}$ is the force induced by the gauge pressure in the chamber, $F_{\text{s}}$ is the force due to capillary forces at the interface, and $F_{\text{v}}$ is the viscous force exerted at the channel walls.

Assuming that the contact line moves at the same velocity as the average liquid height in the channel, the inertial term on the left hand side of \eqref{eq:momentum} can be further expanded as:

 \begin{align}
     \frac{\text{d}}{\text{dt}}\left(m\dot{h}\right) = \frac{\text{d}}{\text{dt}}\left(\rho L \delta h \dot{h}\right) = \rho L \delta \left(\dot{h}^2 + \ddot{h}h\right)\,,
\end{align} where the dots denote differentiation in time.

The first two terms on the right hand side do not require particular assumption expect for the uniformity of the pressure head and the liquid height. These are:
 \begin{alignat}{2}
     &F_g && = -m g  = - \rho \delta L h g \\
     &F_{\Delta p} && = \Delta p(t) A_\text{proj} = \Delta p (t) L \delta\,,\end{alignat} where $A_\text{proj}$ is the channel cross-section and the liquid mass is $m = \rho \delta L h$.

The surface tension contribution $F_\text{s}$ in equation \eqref{eq:momentum}, depends on the shape of the meniscus, which in turn depends on the (unknown) contact angle. Denoting as $f(x)$ the interface shape with respect to the interface location at the center of the channel, any curvature of the interface in the span-wise direction is neglected and the Young-Laplace equation gives the capillary induced pressure difference at the interface:

 \begin{align}
     \Delta p_{\text{s}}  = \frac{\sigma}{\delta} \int_{-\delta/2}^{\delta/2} \kappa(x) \text{d}x, \quad \kappa(x) = \frac{f''}{\left(1 + {f'}^2\right)^\frac{3}{2}}
\end{align} where $\kappa$ is the interface curvature, averaged across the channel width, $f$ denotes the interface shape and the prime denotes spatial derivatives (see also section \ref{sec:lif_setup}. The computation of this term can be done once a model of the interface shape is considered. The model used in this work is described in detail in Section \ref{sec:lif_setup}, and the capillary contribution to equation \eqref{eq:momentum} is $F_{\text{s}}=p_{\text{s}} A_\text{proj}$.

 Finally, the contribution of the viscous forces $F_\text{v}$ in equation \eqref{eq:momentum} depends on the shape of the stream-wise velocity profile near the wall. Assuming that this profile remains parabolic at all times, a shear stress can be estimated: 

 \begin{align}
 \tau &= -\mu\partial_x u=\frac{- 4 \mu u_\text{max}}{\delta} = -\frac{12 \mu}{\delta} \dot{h}
 \end{align} where the maximum velocity is linked to the mean velocity ($\dot{h}$) using the assumption of parabolic velocity profile. The contribution of the wall shear stress in equation \eqref{eq:momentum} is thus $F_\text{visc} = \tau A_\text{wet} = -h \dot{h} {12 L \mu}/{\delta} $, with $A_\text{wet}$ being the wetted surface of the channel walls.

Given a model for the interface shape $f(x)$ (which depends on the evolution of the dynamic contact angle $\Theta(t)$), and given the time-varying pressure in the chamber $\Delta p(t)$, it is possible to integrate equation \eqref{eq:momentum} and compute the evolution of the liquid height under the aforementioned assumption. While the presented model is oversimplified (it does not account, among other things, for the pressure losses at the channel entrance), it is possible to analyze the relative impact of the various terms to the interface dynamics. Moreover, this model was used to link, at least approximately, the pressure in the chamber to the velocity and acceleration of the contact line during the various testing conditions.

 \subsection{Experimental Conditions and Test Cases}\label{sec:Test_Cases}

In the LIF-based flow visualization the experiments were conducted in four conditions, namely three advancing cases and one descending case. \textcolor{black}{The advancing test cases were carried out setting $\Delta p(0)=0$ and $\Delta P$=1000, 1200 and \SI{1500}{\pascal}. These led to $\Delta p(t\rightarrow \infty)$=660, 825 and \SI{985}{\pascal} respectively. The descending case was carried out with $\Delta p(0)=\SI{1200}{\pascal}$ and $\Delta p(t\rightarrow \infty)=0$. The relevant experimental parameters are listed in Table \ref{tab:lif_experiments}. In these experiments, the camera position was not varied: this was placed at the top of the channel, with the upper edge of the image at approximately \SI{15}{\milli\meter} from the upper edge. }

\textcolor{black}{All experiments were repeated three times. Since no appreciable differences were observed in the interface dynamics, we report in this article only the result of a single run per test case. Concerning the descending test case, the choice of presenting only one case is due to the limited impact observed by the initial step $\Delta p(0)$ on the evolution of the interface shape: during the descent, the interface tends to reach a constant velocity after a certain distance and this velocity is fairly independent of the initial pressure step. The chosen value is the one that allows for observing the interface over a large field of view, providing enough space for the video acquisition at multiple heights. }

 \begin{table}[h]
 \centering
 \caption{Experiments in the LIF-based visualization.}
 \vspace{6pt}
 \begin{tabular}{c|cccc}
 \toprule
  LIF & $\Delta p(0)$ & $\Delta P$ & $\Delta p (t \rightarrow \infty)$  & camera pos.\\
 \midrule 
 Test Case 1 & 0 & 1000 & 660& 0\\
 Test Case 2 & 0 & 1200 & 825& 0\\
 Test Case 3 & 0 & 1500 & 985& 0\\
 Test Case 4 & 1200 & - & 0& 0\\
\bottomrule 
 \end{tabular}\label{tab:lif_experiments}
 \end{table}

\begin{table}[h]
 \centering
 \caption{Experiments in the TR-PIV campaign.}
 \vspace{6pt}
 \begin{tabular}{c|cccc}
 \toprule
  TR-PIV & $\Delta p(0)$ & $\Delta P$ & $\Delta p (t \rightarrow \infty)$& camera pos \\
 \midrule 
 Test Case 1 & 0 & 1500 & 985 & 0\\
 Test Case 2 & 200 & 1200 & 985 & 0\\
 Test Case 3  & 1200 & - & 0 & 0\\
 Test Case 4  & 1200 & - & 0 & 40\\
 Test Case 5  & 1200 & - & 0 & 120\\
 \end{tabular}\label{tab:piv_experiments}
 \end{table}

  For the PIV measurements, two rising cases and three descending cases were considered. \textcolor{black}{The experimental parameters are listed in Table \ref{tab:piv_experiments}. The `Test Case 1' in the PIV campaign corresponds to the conditions of 'Test Case 3' in the LIF-visualization campaign. In the Test Case 2, the acceleration was reduced by reducing the pressure step setting $\Delta p(0)=200$ Pa. The other test cases are in the same conditions as `Test Case 4' in the LIF-visualization campaign, and the only difference is in the position of the camera which is moved downward by \SI{40}{\milli\meter} and \SI{120}{\milli\meter} with respect to the reference value. As for the LIF visualization experiments, also the TR-PIV experiments were repeated three times and no significant variability was observed.}



 \section{Measurement Techniques}

 \subsection{LIF Visualization: Set-up and Processing} \label{sec:lif_setup}
 
For the LIF measurements, Rhodamine B from Sigma Aldrich is dissolved in the water. Images were captured with a Phantom v2012 high-speed camera which has a resolution of 1200$\;\times\;$800 px. The particles were illuminated with a continuous Stabilite 2017 ion laser system with a wavelength of \SI{515}{\nano \meter}. The power of the laser was set to \SI{1.9}{\watt}. An objective lens with a focal length of \SI{105}{\milli\meter} was used. The camera was placed at a larger distance from the channel compared to the PIV experiments, to allow the tracking of the interface over a large vertical distance. The achieved optical magnification was 30 px/mm. Images were acquired in single frame mode, with a frame rate of \SI{500}{\hertz} and an exposure time of \SI{20}{\micro\second}. In total, 1500 images were acquired, resulting in a measurement duration of \SI{3}{\second}.

 \begin{figure}[htp]
     \centering
     \subfloat[]{\label{fig:before_prep}
       \includegraphics[width=.23\textwidth]{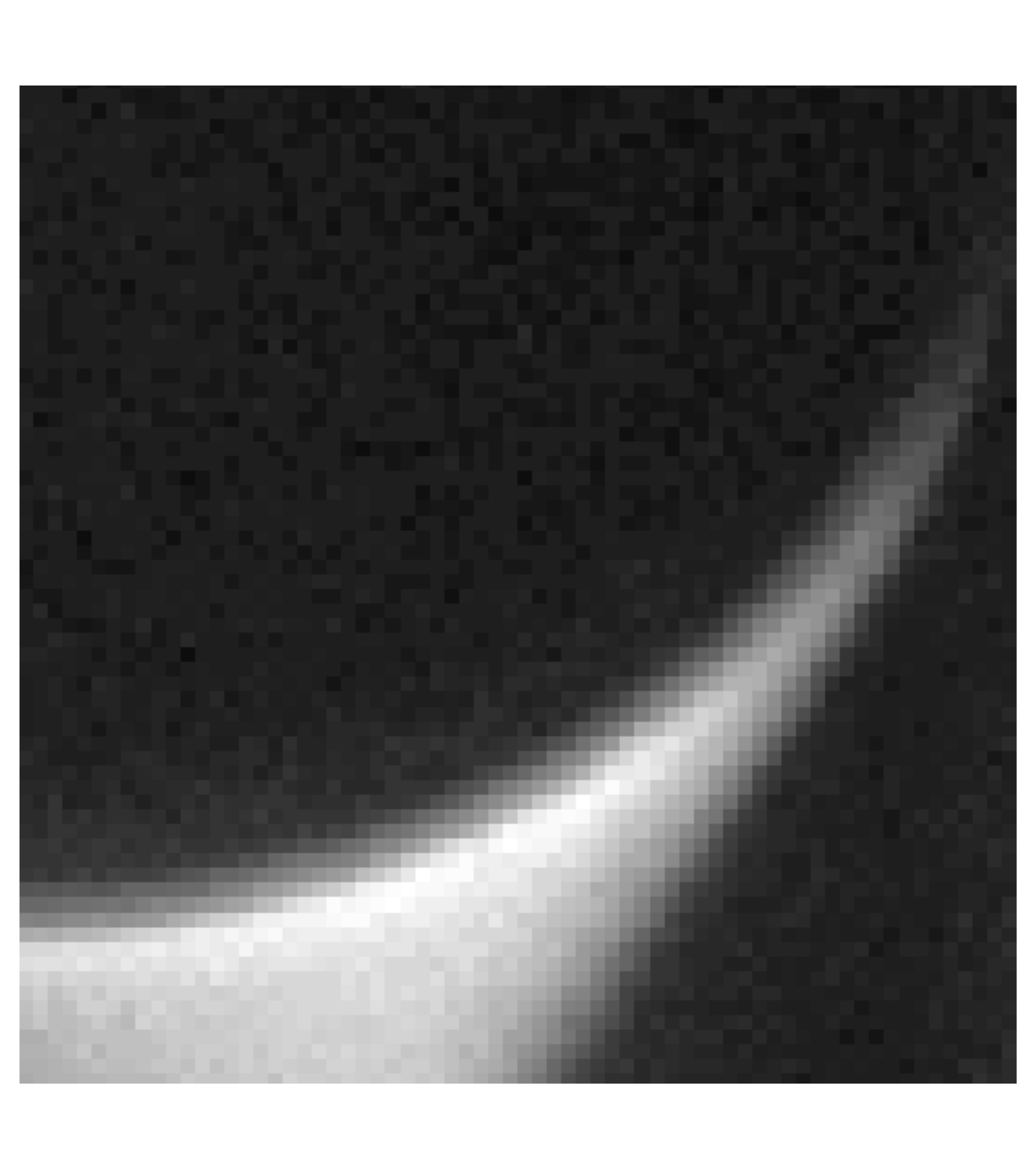}}
 \subfloat[]{\label{fig:after_prep}
       \includegraphics[width=.23
       \textwidth]{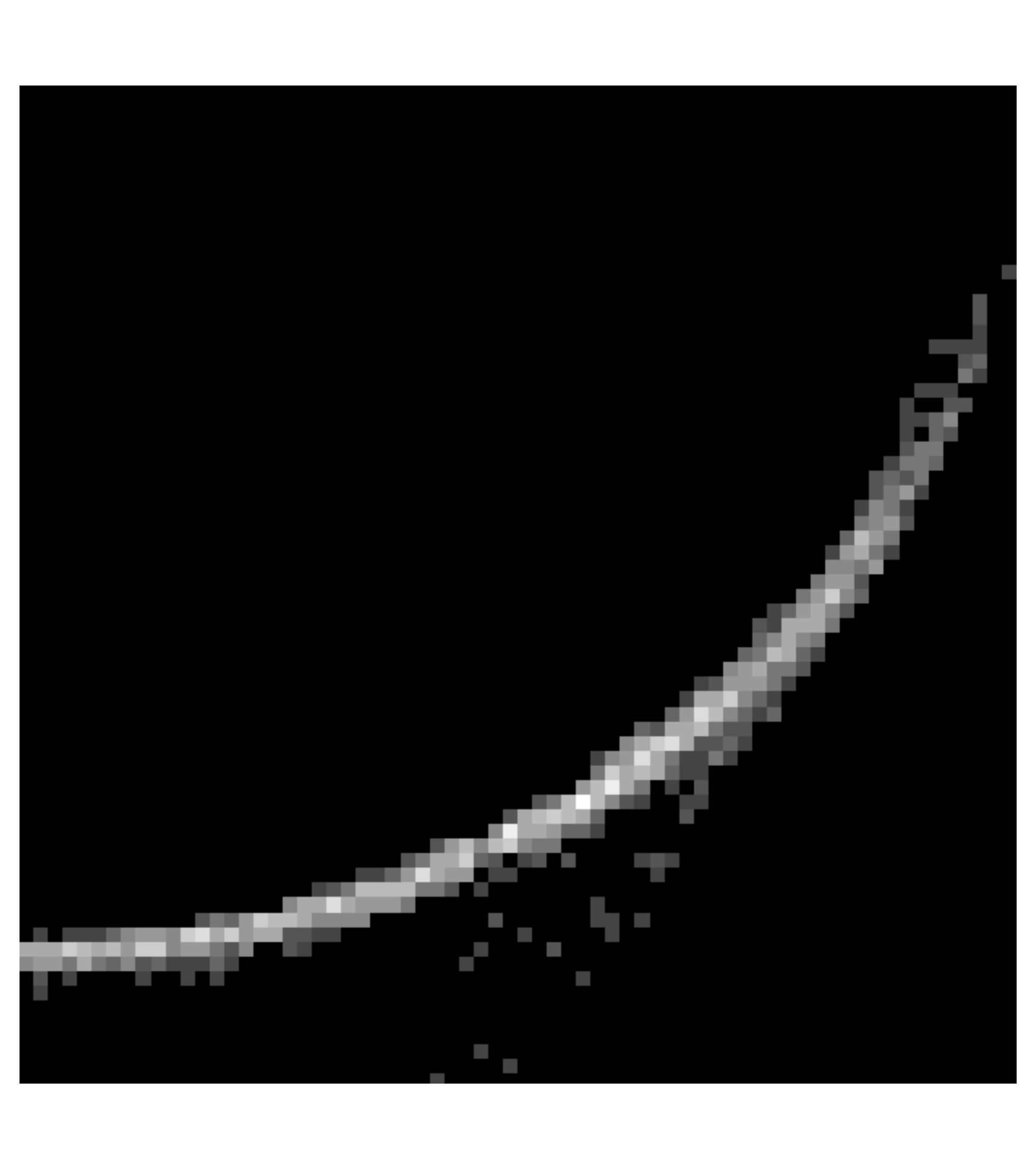}}

     \caption{Fig (a) Raw image of the meniscus obtained from the experiments. Fig (b) Meniscus after non-local means denoising and high-pass filtering. }\label{fig:raw_menisci}
\end{figure}

 A zoom view of the LIF-based visualization of the dynamic meniscus is shown in Figure \ref{fig:raw_menisci}. The images were processed using the non-local means denoising by \cite{cv2} and then high-pass filtered to highlight the gas-liquid interface. The result of this step, for the image in Figure \ref{fig:before_prep}, is shown in Figure \ref{fig:after_prep}. Only the right side of the symmetric meniscus is shown. The interface is detected by calculating the vertical gradient of the image intensity, similarly to \cite{Mendez2016,Mendez2018}, by searching for the peak in the intensity gradient in each image column. At the end of this process, outliers were removed using a local median filter. 

\begin{figure}[htp]
     \centering
   \includegraphics[width=.24\textwidth]{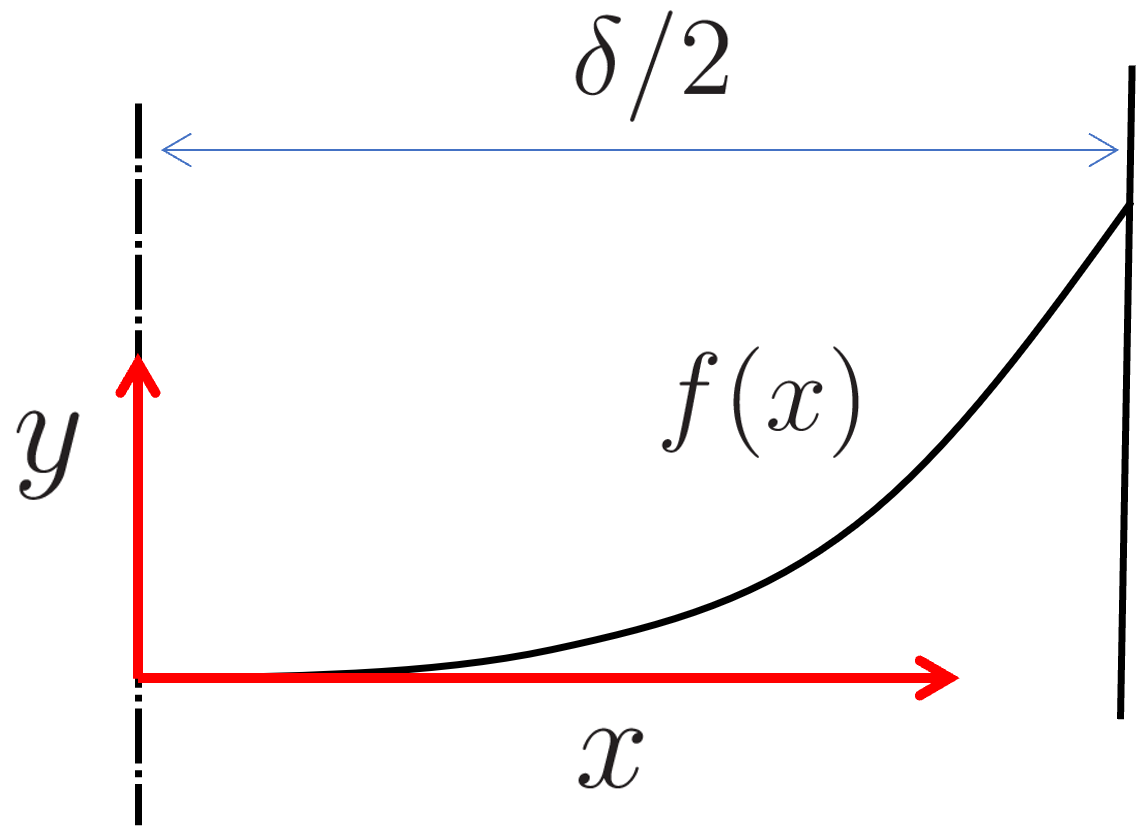}
     \caption{Location of the reference frame moving with the meniscus, and with respect to which the model in equation \eqref{cosh} is defined.}\label{Fig_M}
\end{figure}

 A simple model for the interface shape is then fitted to the detected points. This model was heuristically adapted from the analytical solution of a meniscus in stationary conditions and reads: 

 \begin{equation}
 \label{cosh}
     f(x) = \text{cosh}\left(\frac{x^a}{b}\right)-1,
 \end{equation}where $y=f(x)$ is the vertical position of the interface with respect to the interface at the center of the image (see Figure \ref{Fig_M}) and $a, b \in \mathbb{R}$ are constants to be fit at each time step.

 The fitting is performed using the \textsc{minimize} function from Python's Scipy library \citep{2020SciPy-NMeth} to minimize the $l_2$ norm of the prediction error and the time evolution of each parameter is further smoothed using a low-pass filter to eliminate outliers.

 Equation \ref{cosh} is used to (1) compute the contact angle by solving $\text{ctg}(\Theta)=f'(\delta/2)$, with $'$ denoting the spatial derivative, (2) to compute the location of the contact line as $h(\delta/2,t)=f(\delta/2)+h(t)$, with $h(t)$ the liquid height at the center of the channel and (3) compute the contact line velocity as $u_c(t)=\dot{h}(\delta/2,t)$.

\subsection{TR-PIV: Set-up}\label{sec:piv_setup}

For the PIV measurements, the fluid was seeded with \textit{Red Fluorescent Polymer Microspheres} from {Thermo Fisher Scientific} with a diameter of \SI{12}{\micro\meter}. The flow was observed with a TR-PIV system from Dantec Dynamics. The particles were illuminated with the laser (DM40-527-DH Nd:YLF from Photonics Industries) with a wavelength of \SI{527}{\nano\meter} and a maximum pulse energy of \SI{40}{\milli\joule} at a frequency of \SI{1}{\kilo\hertz}. Images were recorded with a SpeedSense Ethernet M310 camera having a resolution of 1280$\times$800 px (at a maximum frame rate of \SI{3260}{\hertz}). \textcolor{black}{The acquisition frequency of both laser laser and camera, operating in \emph{single frame mode}, was and \SI{1.2}{\kilo\hertz}}. An objective lens with a focal length of \SI{105}{\milli\meter} was used to get an optical magnification of 60 px/mm.

\begin{figure}[htp]
     \centering
    \includegraphics[width=.47\textwidth]{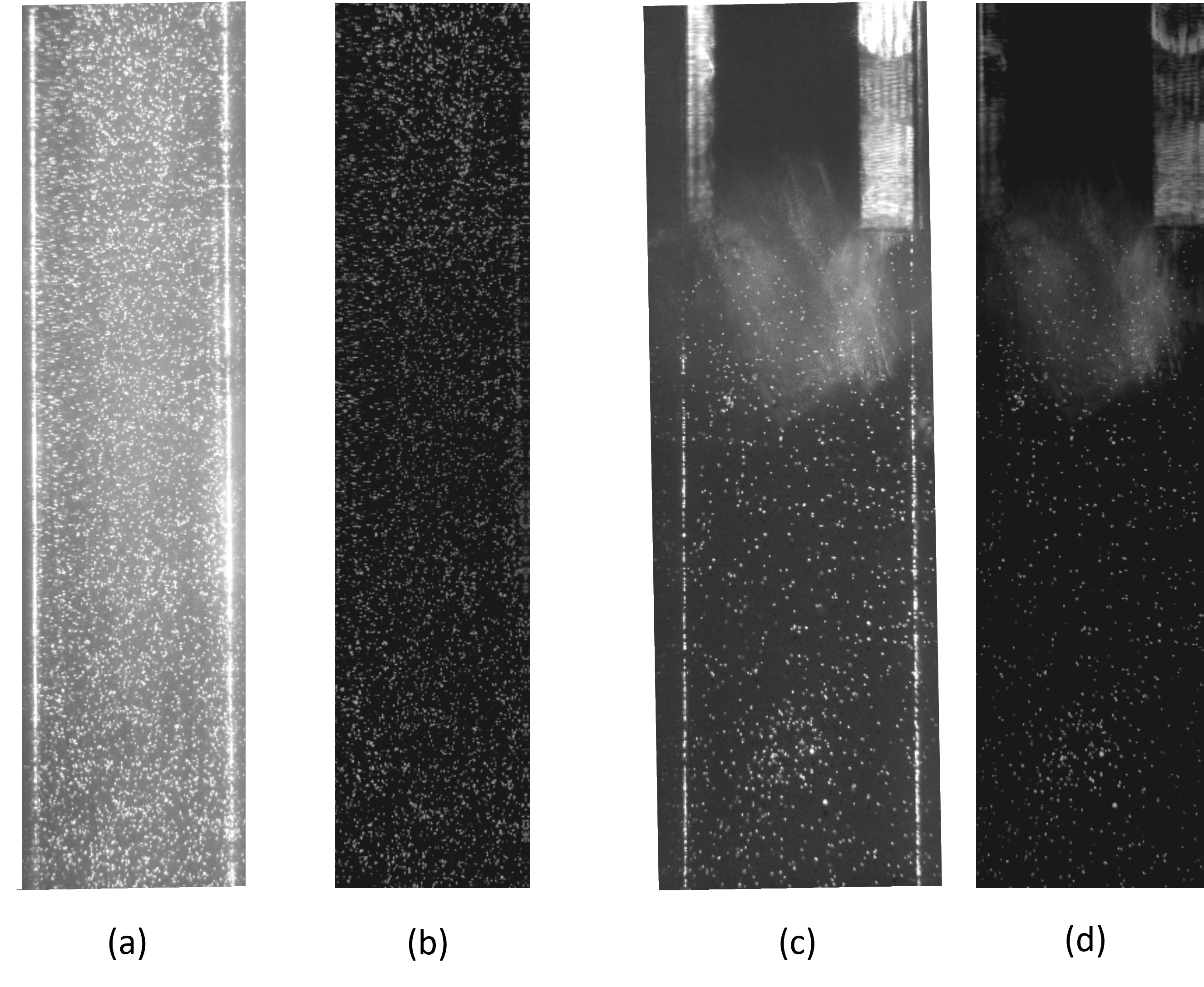}
     \caption{PIV snapshots before and after the pre-processing. The interface is not in the field of view in figure (a) while it is in the field of view in (c). Figures (b) and (d) result from the eigenbackground removal from (a) and (c) respectively.}\label{PIV_FIG}
 \end{figure}

Figure \ref{PIV_FIG} shows two snapshots from the PIV acquisition \emph{before} (a,c) and \emph{after} (b,d) the image pre-processing, which was carried out using the POD-based background removal by \cite{mendez2017}.
 Figure \textcolor{red}{6a} shows a snapshot in which the interface is not within the field of view; in this case the illumination and the seeding was fairly homogeneous. Figure \textcolor{red}{6c} shows a snapshot in which the interface is in the field of view; in this case the interface shape and the light refraction hinder the interface detection and the velocimetry in its proximity.

 This is why the PIV interrogation was carried out in a cropped region of the FOV moving with the interface and having its upper boundary (edge 4-3 in Figure \ref{fig:exp_theory}) approximately \SI{1.5}{\milli\meter} below the expected liquid level. This distance was measured in several of the snapshots for which the interface is visible and was kept constant by a tracking algorithm. 
 In particular, the Region of Interest (ROI) in which the PIV was carried out, is displaced for each image pair $k$, at time $t_k$, using the mean velocity computed along the bottom edge 1-2 (denoted as $y_{1,2}$), i.e.  $\overline{v}_k=\int_{-\delta/2}^{\delta/2} v(x, y_{1,2},t_k) dx$. The vertical displacement of the FOV is thus $
 \Delta x=\overline{v}\Delta t$, with $\Delta t$ the sampling frequency of the PIV. The velocity of the ROI was found to closely match the mean velocity of the interface $h(t)$.

In the ROI, the PIV interrogation was carried out with the \textit{OpenPIV} software by \cite{openpiv}, using windows of large aspect ratio, justified by the fact that velocity component $v$ (along $y$) is much larger than the velocity component $u$ (along $x$). The rectangular interrogation windows allows for a better sampling of the flow in the direction where the largest gradients are expected, without deteriorating the signal-to-noise ratio in the cross-correlation maps.

 Adaptive iterative multigrid interrogation \citep{scarano2000} was used to calculate the velocity fields. For the advancing contact line experiments, the initial window size is set to 256$\times 64$ px and reduced over two steps to 64$\times 16 $ px. For the receding contact line experiments, the initial size is 64$\times$64 px, and reduced to 24$\times$24 px in two iterations. Outliers were removed based on the signal-to-noise ratio measured in terms of peak to standard deviation in the cross-correlation map.

 \subsection{Superresolution of PIV fields}\label{sec:piv_sup}

 \textcolor{black}{ The post-processing of the PIV fields aimed at increasing the resolution of the PIV data and enable accurate computation of $Q-$fields. The approach consist in using a Radial Basis Function (RBFs, see \cite{fornberg2015}) expansion of the velocity field to compute derivatives analytically. A similar approach has been used for robust interpolation \citep{Casa2013}, to compute derivatives near walls \citep{kkarri2009} and compute pressure fields from PIV \citep{sperotto2021}. In this work, we use it to compute the $Q-$fields analytically and thus accurately measure the intensity of the vortices near the gas-liquid interface.}
 
  \textcolor{black}{ 
 To reduce the computational cost of the RBF regression, we combine it with a classic Proper Orthogonal Decomposition (POD). The main idea behind the POD-RBF-super-resolution strategy is to perform RBF regression of the spatial and temporal structures of the decomposition, then use these to reconstruct high-resolution POD modes and finally rebuild high-resolution velocity fields. The procedure is herein briefly summarized. Let}

 \begin{equation}
 \label{POD_T}
     \vec{u}(\textbf{x}_0, t_0) = \sum_{i = 1}^R \sigma_i \vec{\phi_i}(\textbf{x}_0) \varphi_i(t_0)\,,
 \end{equation} \textcolor{black}{denote the truncated POD of the velocity field on an initial spatial grid 
  $\mathbf{x}_0$ containing $n_s$ points and a temporal grid $t_0 $ with $n_t$ points. Here $R\ll n_t$ denotes the truncation index, $\sigma_i$ is the amplitude of the i-th POD mode with spatial structure $\vec{\phi_i}(\textbf{x}_0)$ and temporal structure $\varphi_i(t_0)$.} 
  
  The details of the POD computation can be found elsewhere (e.g. \cite{Mendez2019,Mendez2020}) and are omitted here.   \textcolor{black}{We denote as } 
  \begin{equation}\label{EQ1}
     \vec{\phi}_i(\textbf{x}_0) = \sum_{j = 1}^{n_\phi} w^{\phi}_{ij} \,\gamma^{\phi}_{j} (\mathbf{x}_0| \mathbf{x}_{i,j},\Sigma_{\phi_j}),,  
 \end{equation} and \begin{equation}\label{EQ2}
     {\varphi}_i(t_0) = \sum_{j = 1}^{n_\varphi} w^{\varphi}_{i,j} \,\gamma^{\varphi}_j (t_0| t_j,\Sigma_{\varphi_j})\,
\end{equation} the RBF expansion of the spatial and temporal structures respectively.
Therefore, $w^{\phi}_{i,j}$ and $w^{\varphi}_{i,j}$ are the set of weights defining the regression functions, $\gamma^{\phi}_j (\mathbf{x}_0| \mathbf{x}_j,\Sigma_{\phi_j})$  are the $n_\phi$ radial basis functions in space and $\gamma^{\varphi}_j (t_0| t_j,\Sigma_{\varphi_j})$ are the $n_\varphi$ radial basis functions in time. These have collocation points $\mathbf{x}_j$ and $t_j$ and shape parameters $\Sigma_{\phi_j}$ and $\Sigma_{\varphi_j}$.  Note that the coefficients $w^{\phi}_{i,j}$ must be defined as vectors, i.e. $w^{\phi}_{i,j}=(w^{\phi}_{u,i,j},w^{\phi}_{v,i,j})$ since the same radial basis $\gamma^{\phi}_j$ is used for both components $\vec{\phi}=(\phi_u,\phi_v)$. However, because every snapshot is reshaped as a column vector, regardless of whether this collects a vector or a scalar field, we keep the same notation as for the interpolation of the temporal structures. \textcolor{black}{We consider Gaussian RBFs both in space and in time. The spatial RBFs $\gamma^{\phi}_j (\mathbf{x}_0| \mathbf{x}_j,\Sigma_{\phi_j})$ are defined as }

 \begin{equation}
 \label{eq:Gaussian_RBF}
     \gamma^{\phi}_j= \exp\left(-(\mathbf{x}_0-\mathbf{x}_j)^T\Sigma^{-1}_{\phi_j}(\mathbf{x}_0-\mathbf{x}_j)\right)
  \end{equation} with $\Sigma^{-1}_{\phi_j}=\text{diag}(1/2\sigma^2_x,1/2\sigma^2_y)$, while the temporal RBFs $ \gamma^{\varphi}_j (t_0| t_j,\Sigma_{\varphi_j})$ are defined as
\begin{equation}
     \gamma^{\varphi}_j= \mbox{exp}\left(-(t_0-t_j)^2 / (2 \sigma^2_t) \right)\,
 \end{equation} with $\Sigma_{\varphi_j}=1/2\sigma^2_t$.
 
  For both space and time regressions, the collocation point and the shape parameters are defined a priori and the regression is solved once the weights are identified. Reshaping all bases functions as columns of matrices $\Gamma_{\phi}(\mathbf{x}_0)\in\mathbb{R}^{n_s\times n_\phi}$ and $\Gamma_{\varphi}(t_0)\in\mathbb{R}^{n_t\times n_{\varphi}}$, collecting all the unknown weights into column vectors $\mathbf{w}^{\phi}\in\mathbb{R}^{n_\phi}$, $\mathbf{w}^{\varphi}\in\mathbb{R}^{n_\phi}$, and collecting all the entries of $\vec{\phi}_i(\textbf{x}_0)$ and ${\varphi}_i(t_0)$ into column vectors $\mathbf{\phi}_0\in\mathbb{R}^{n_s}$ and  $\mathbf{\varphi}_0\in\mathbb{R}^{n_t}$, the weights are the solution of classic least square problems. Using a Tikhonov regularization to mitigate the risks of over-fitting \citep{Mendez2020_LS}, the solution is:

 \begin{equation}
     \mathbf{A} \mathbf{x}=\mathbf{b} \quad \rightarrow \mathbf{x}=\Bigl(\mathbf{A}^T\mathbf{A}+\alpha\mathbf{I}\Bigr)^{-1} \mathbf{A}^T \mathbf{b},
   \label{eq:inverseproblem}
 \end{equation} where $\mathbf{A}=\Gamma_{\phi}(\mathbf{x}_0)$, $\mathbf{x}=\mathbf{w}^{\phi}$ and $\mathbf{b}=\mathbf{\phi}_i$ for the regression in space and
 $\mathbf{A}=\Gamma_{\varphi}(t_0)$, $\mathbf{x}=\mathbf{w}^{\varphi}$ and
 $\mathbf{b}=\mathbf{\varphi}_i$ for the regression in time. The regularization parameter $\alpha$ is taken as $\epsilon\, ||\mathbf{A}||_F$, with $\epsilon=1e-11$ and $||\,||_F$ the Frobenious norm. Note that the inverse in \eqref{eq:inverseproblem} can be pre-computed and the system solved for all the modes in one single step.

Given the weights, equations \eqref{EQ1} and \eqref{EQ2} can be used on \emph{any} arbitrary mesh grid $\mathbf{x}$ and \emph{any} time discretization $t$. Then, the resulting high resolution mode can be introduced in the expansion \eqref{POD_T}. 

The analytic description of the flow allows for the computation of derivatives by simply replacing the RBF with the required (analytic) derivatives. For example, $\partial_x u$ can be computed by replacing $\vec{\phi}_{i}(\textbf{x})$ with $\partial_x\vec{\phi}_{i}(\textbf{x})$ in \eqref{POD_T} and this can be computed by replacing $ \gamma^{\phi}_j (\mathbf{x}| \mathbf{x}_j,\Sigma_{\phi_j})$ with $\partial_x \gamma^{\phi}_j (\mathbf{x}| \mathbf{x}_j,\Sigma_{\phi_j})$ in \eqref{EQ1}. These derivatives are analytically available because $\partial_x \gamma^{\phi}_j$ is analytically available by differentiating \eqref{eq:Gaussian_RBF}. 

 Finally, using the available derivatives it is possible to compute the strength of vortices in the PIV fields in terms of $Q$-field, which for a 2D incompressible flow reads \citep{hunt1988} 

 \begin{equation}
 \begin{split}
     Q &= \frac{1}{2}(||\Omega||_F^2 - ||S||_F^2)
   \\& = -\frac{1}{2} \left(\left({\partial_x u}\right)^2 + 2{\partial_y u}\,\partial_x v + \left({\partial_y v}\right)^2 \right)
       \end{split}\,
 \end{equation} where $S=1/2( \nabla V+\nabla V^T)$ and $\Omega=1/2( \nabla V-\nabla V^T)$ the symmetric and anti symmetric portions of the velocity gradient tensor.

 \begin{figure*}[h]
 \captionsetup[subfigure]{justification=centering}
     \centering
         \subfloat[]{\label{fig:rise_interface_1}{
         \includegraphics[width=.29\textwidth ]{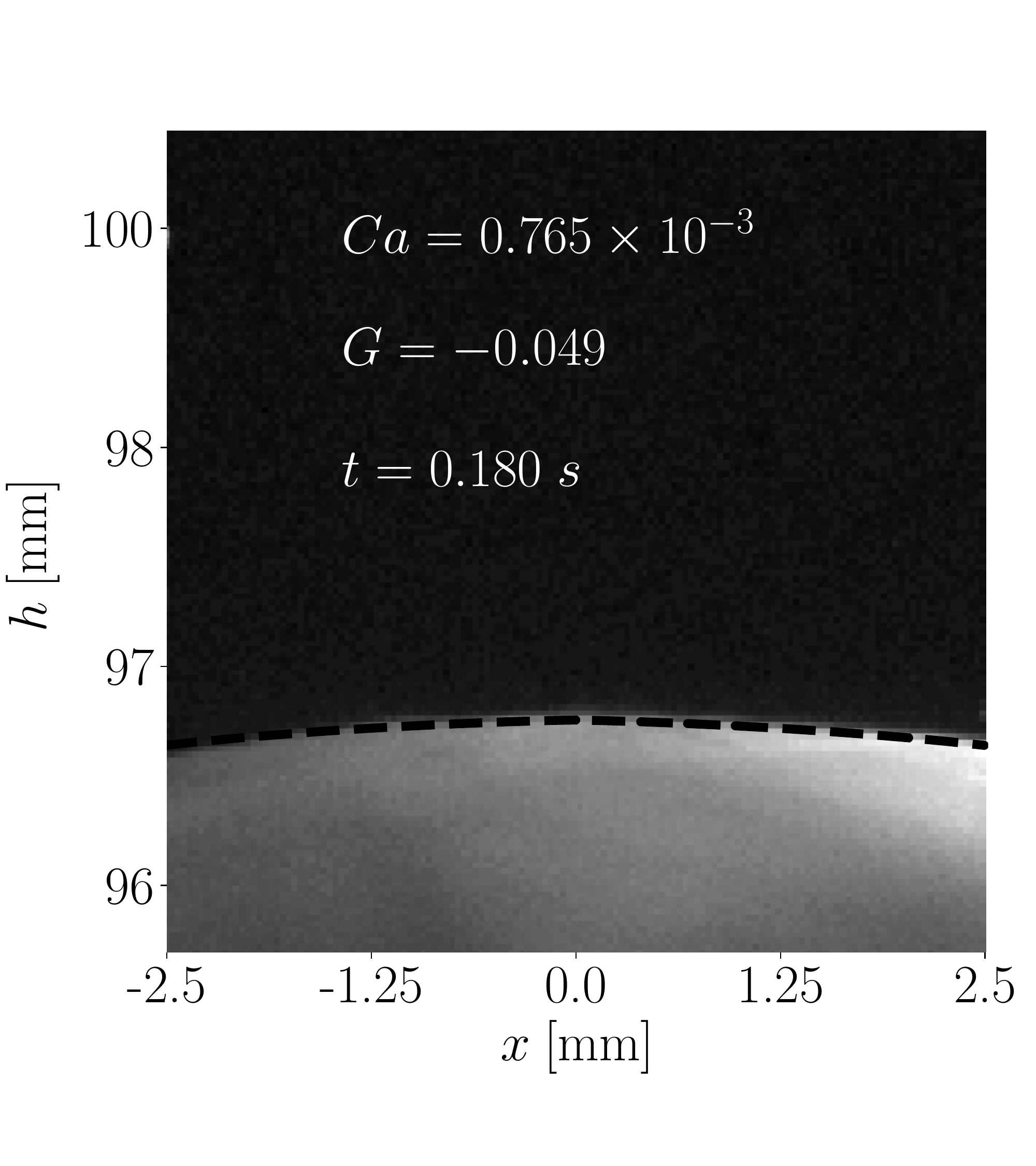}}
 	} \hspace{0.1cm}
 ~
         \subfloat[]{\label{fig:rise_interface_2}{
        \includegraphics[width=.29\textwidth ]{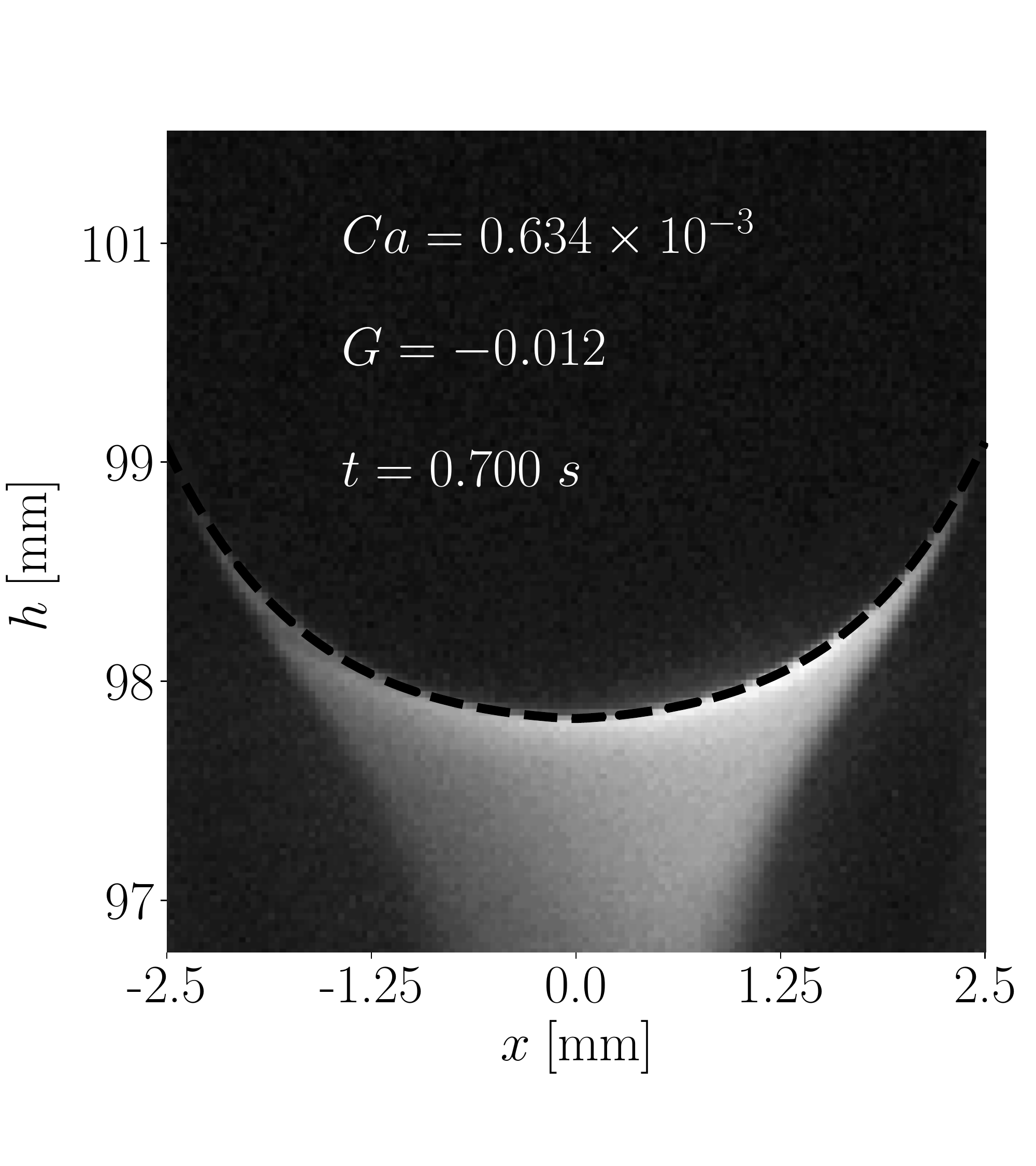}}
 	} \hspace{0.1cm}
 ~
         \subfloat[]{\label{fig:rise_interface_3}{
        \includegraphics[width=.29\textwidth ]{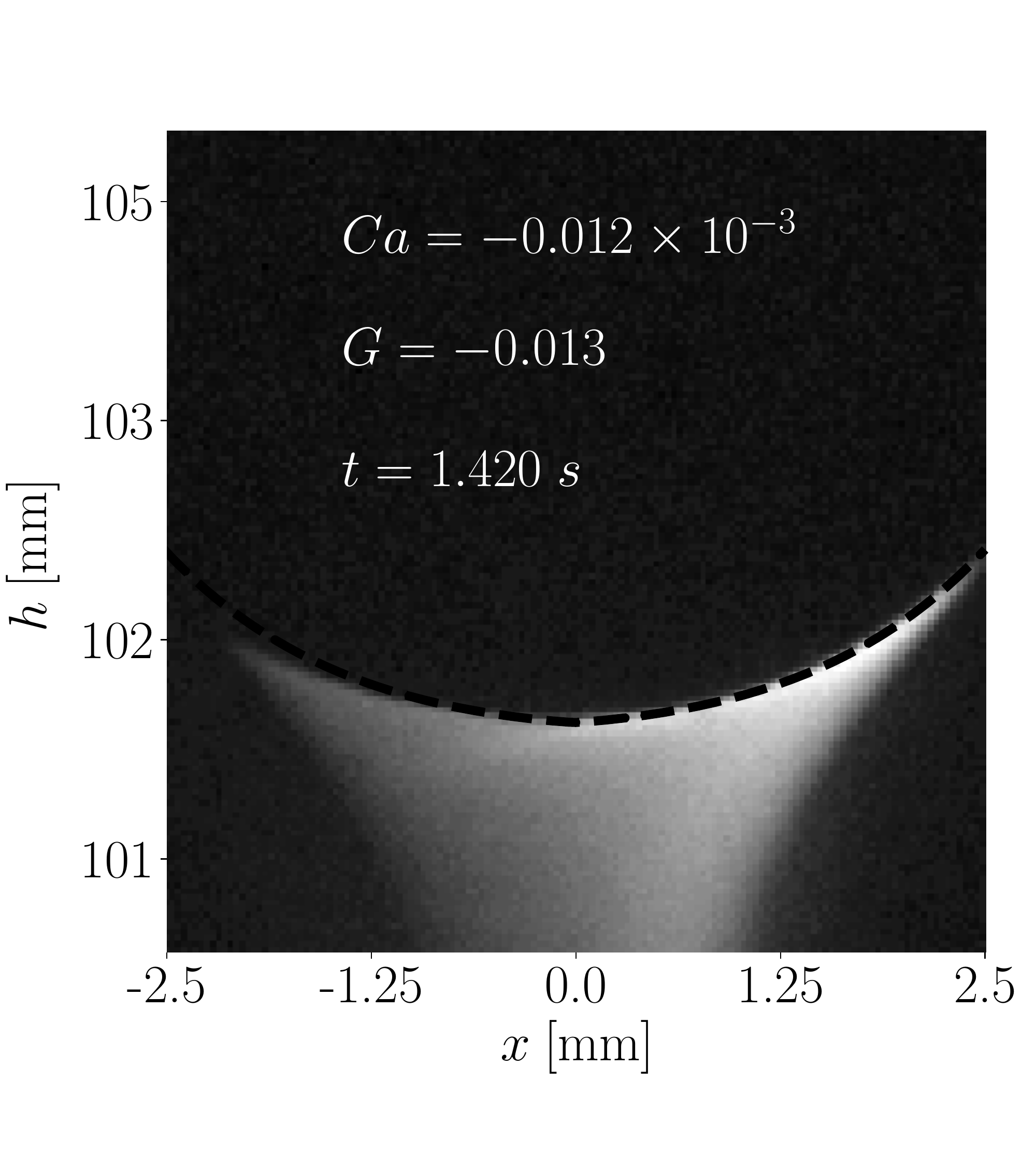}}
 	}
     \caption{LIF results for Test Case 1. Fig (a) corresponds to the interface above the velocity field in Figure \ref{fig:rise_field_2}. The three images are marked by the three red squares in Figure \ref{fig:model_prediction_comparison_1500}. The titles recall the $Ca$ and $G$ numbers for each snapshot as well as the time $t$ at which the meniscus came into the FOV of the camera.}\label{fig:rise_interface}
 \end{figure*}

 \begin{figure*}[h]
     \centering
      \subfloat[]{\label{fig:fall_interface_1}{
    	\includegraphics[width=.29\textwidth ]{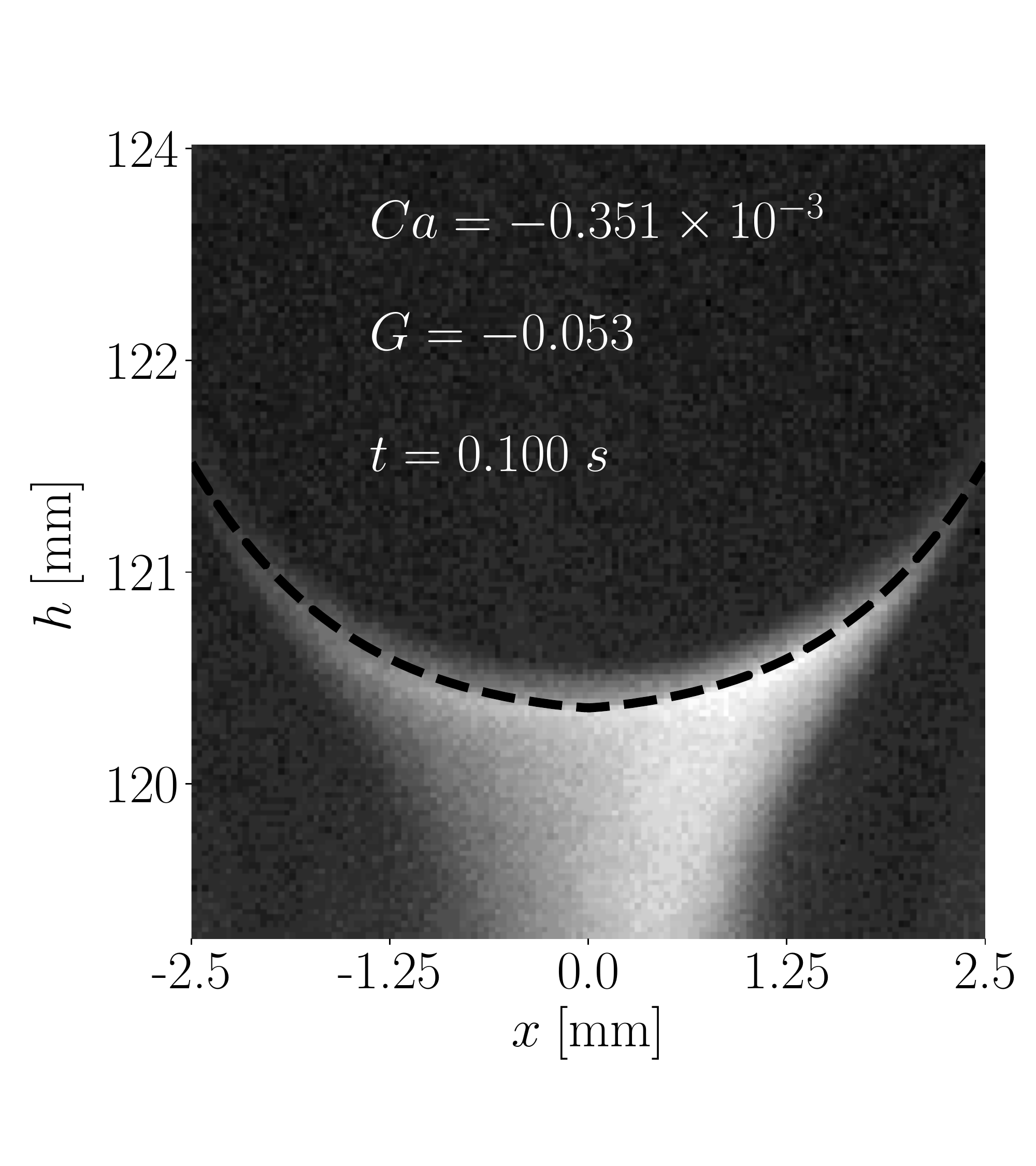}}
 	} \hspace{0.1cm}
 ~
     	 \subfloat[]{\label{fig:fall_interface_2}{
        \includegraphics[width=.29\textwidth ]{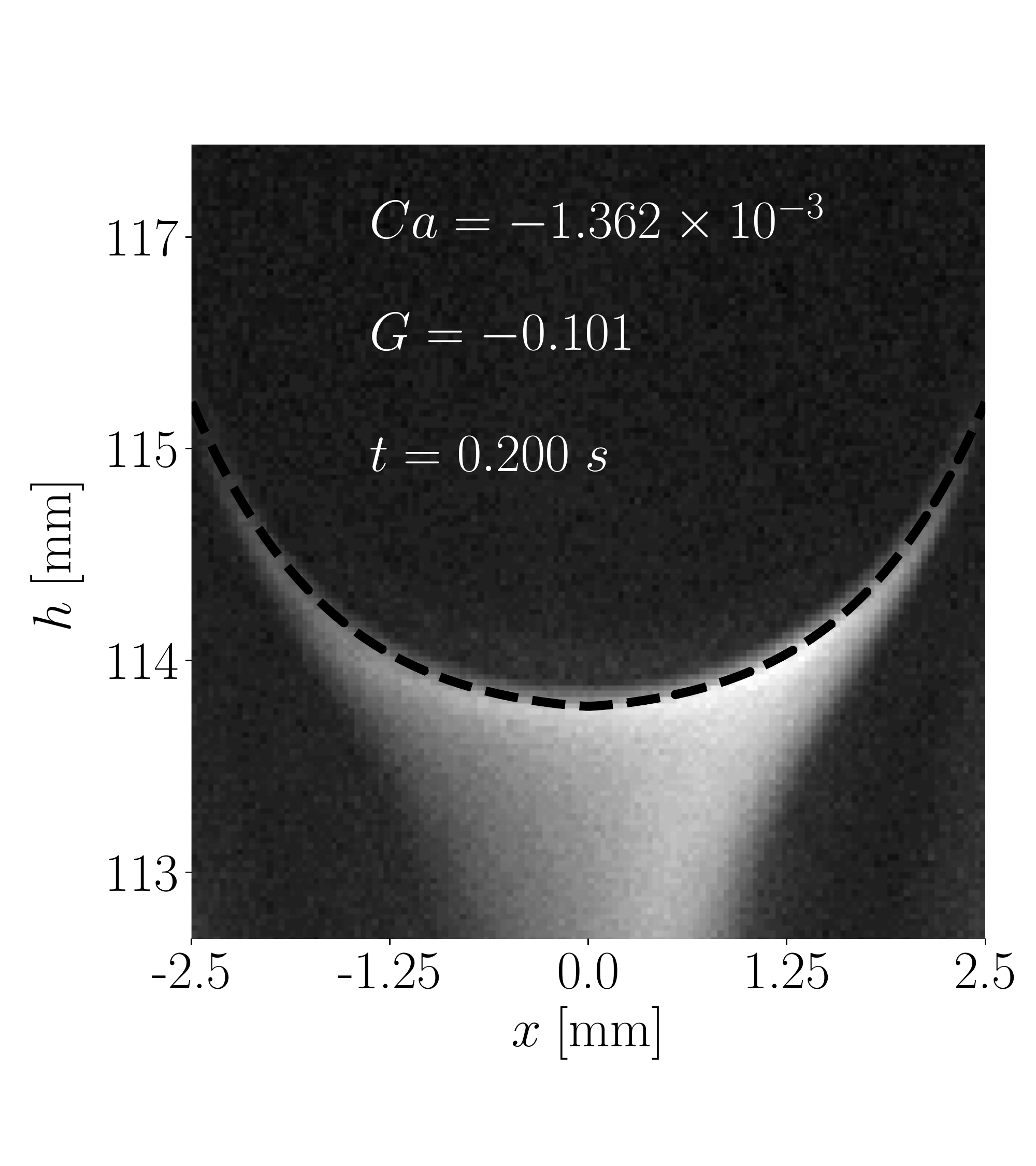}}
 	} \hspace{0.1cm}
 ~
     	 \subfloat[]{\label{fig:fall_interface_3}{
       \includegraphics[width=.29\textwidth ]{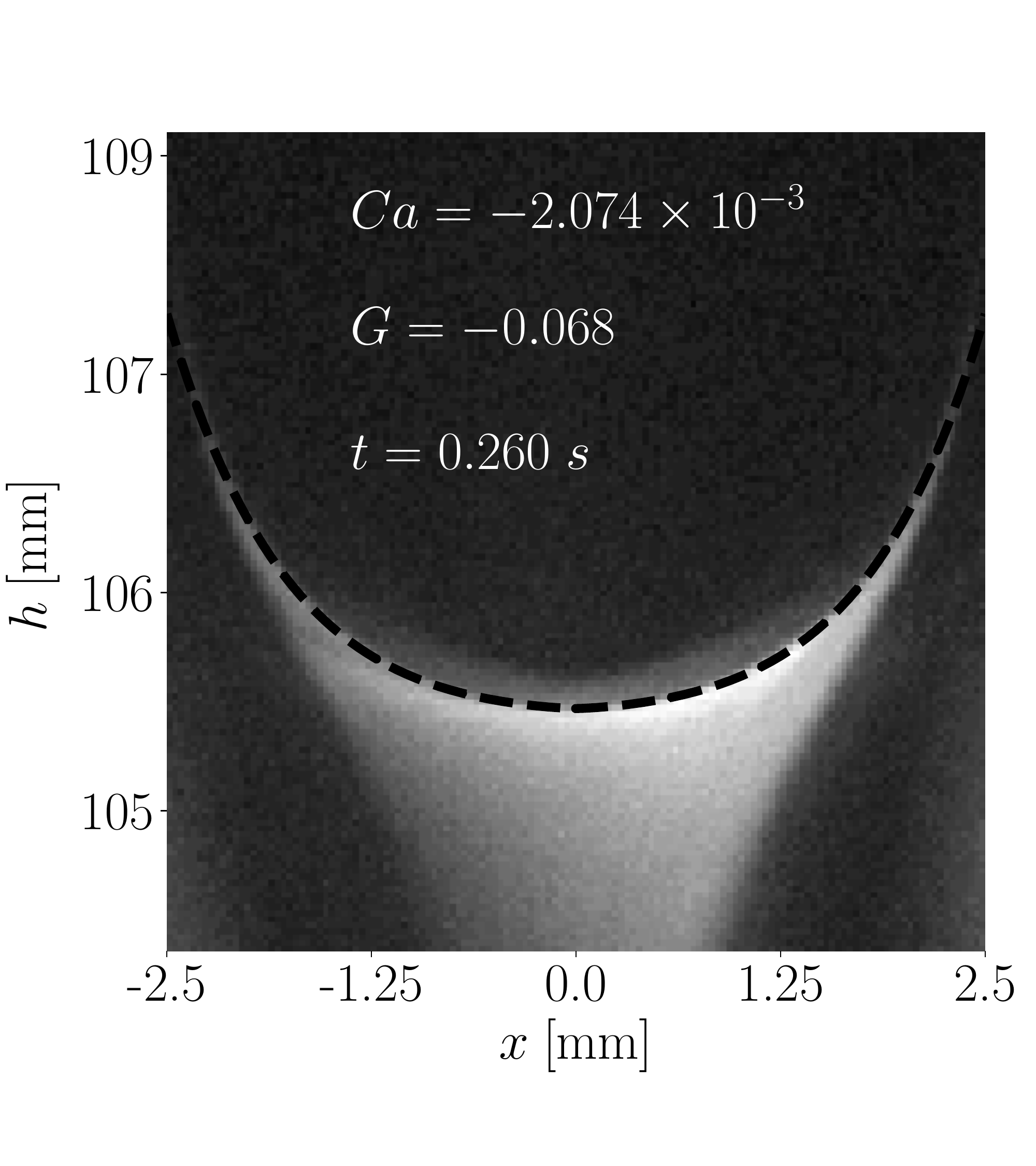}}
 	}
     \caption{LIF results for Test Case 4. In each snapshot, the values of $Ca$ and $G$ are given as well as the time $t$ at which the meniscus came into the FOV of the camera.}\label{fig:falling_interface}
\end{figure*}

 \section{Results} \label{sec:results}

 \subsection{LIF Visualization Results}
 \label{sec:lif_results}

\begin{figure*}[htp]
     \centering
     \subfloat[]{\label{fig:model_prediction_comparison_1000}
       \includegraphics[width=.47\textwidth]{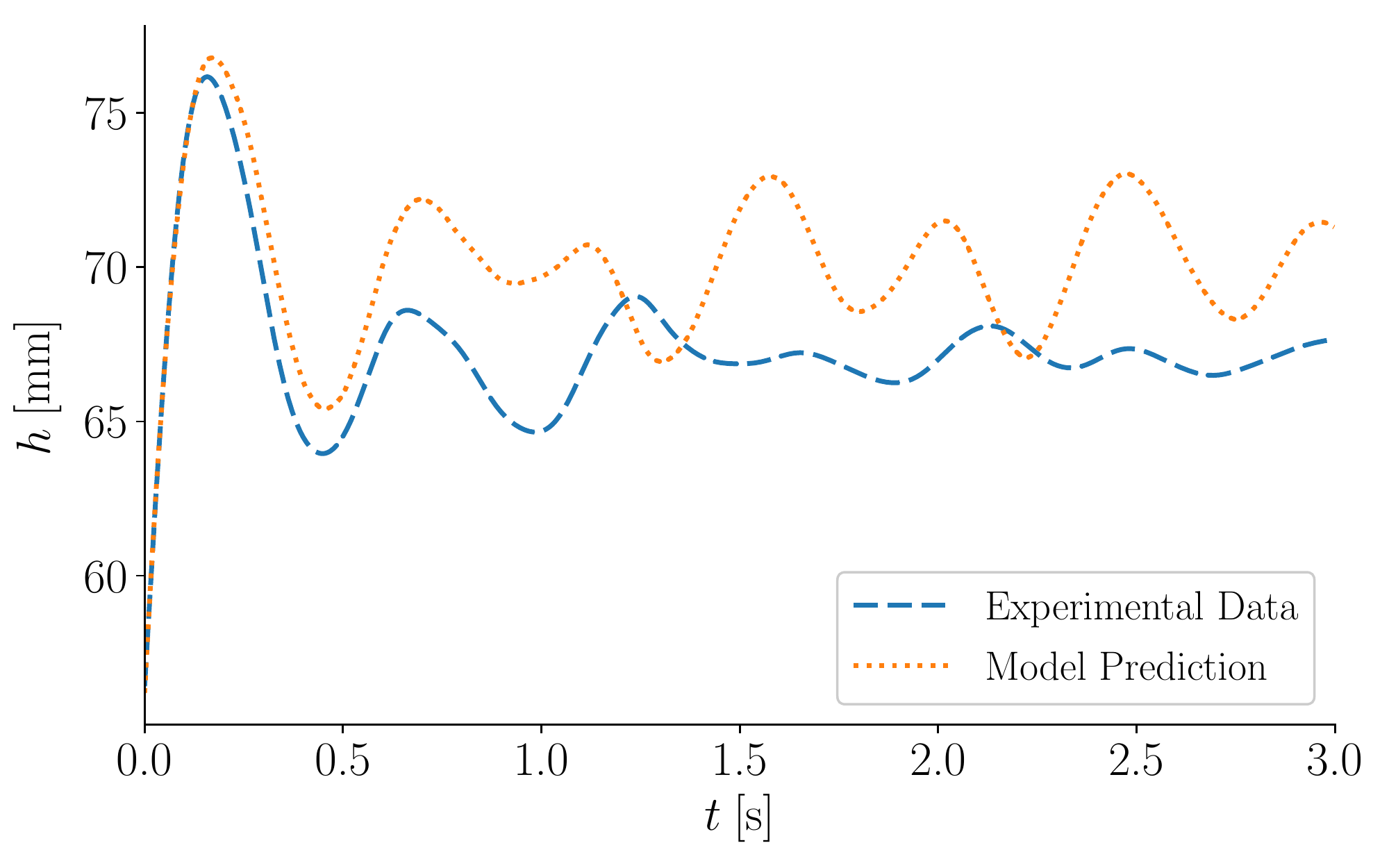}}
     \subfloat[]{\label{fig:theta_over_ca_1000}
       \includegraphics[width=.47\textwidth]{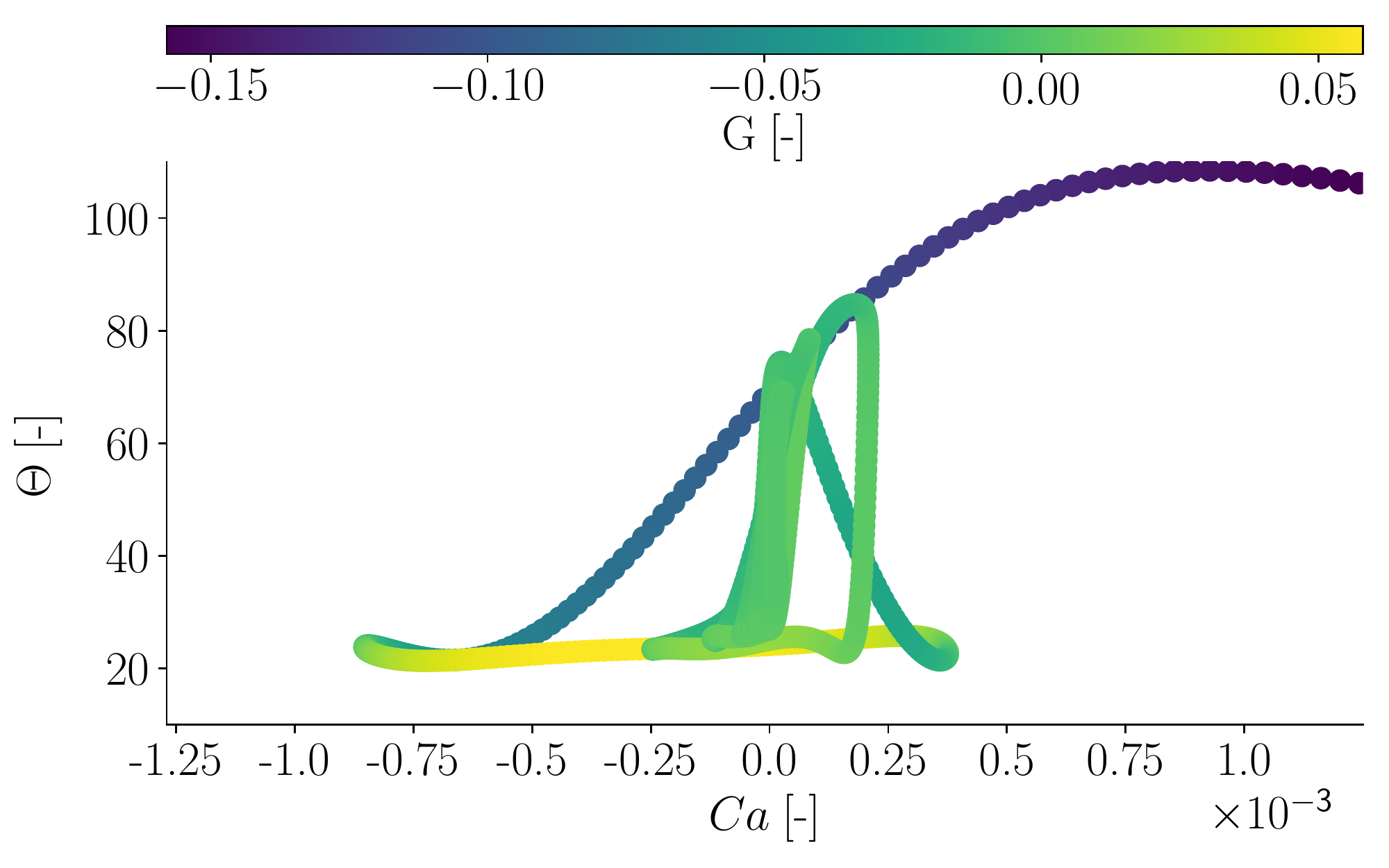}}

     \subfloat[]{\label{fig:model_prediction_comparison_1250}
      \includegraphics[width=.47\textwidth]{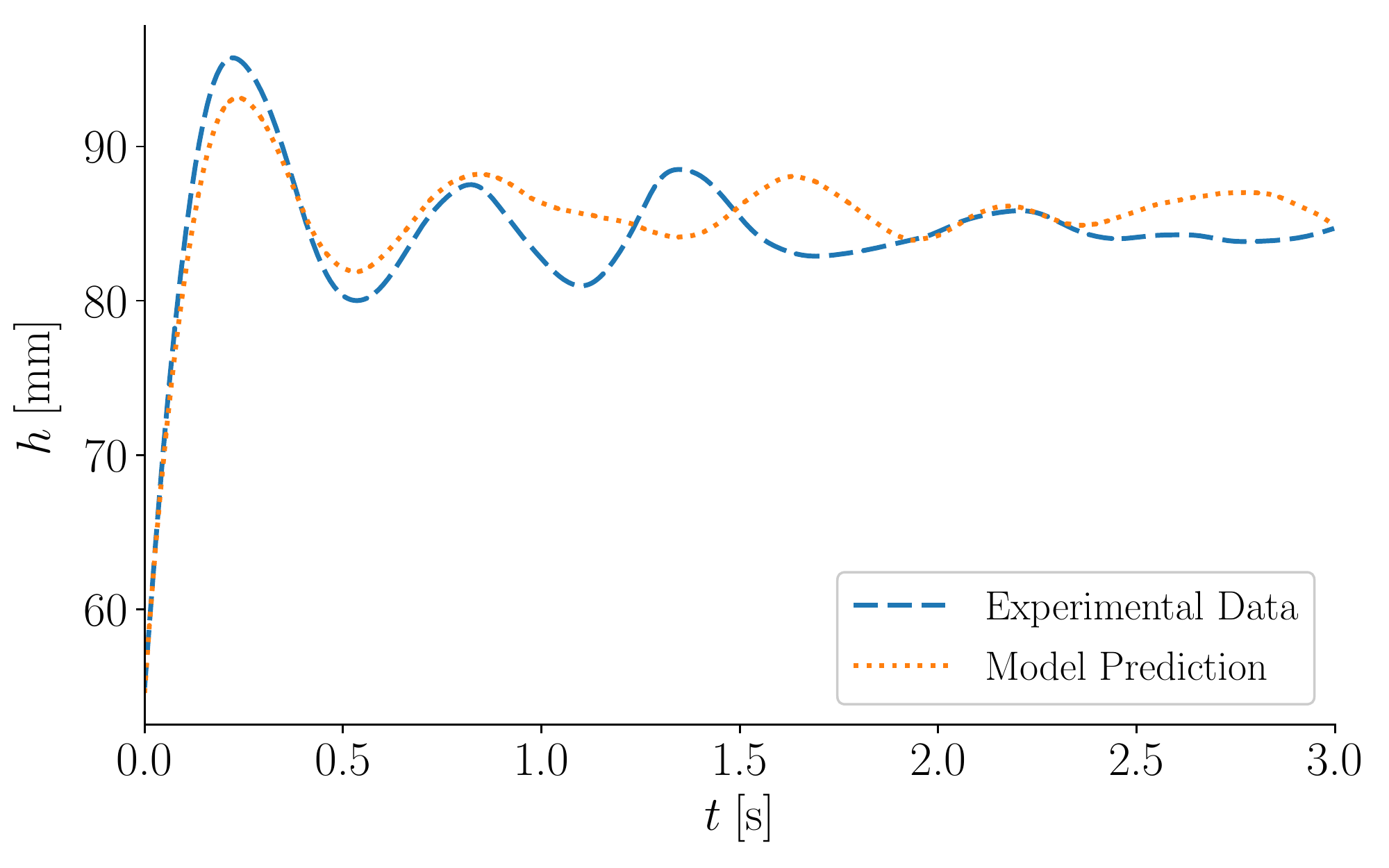}}
     \subfloat[]{\label{fig:theta_over_ca_1250}
       \includegraphics[width=.47\textwidth]{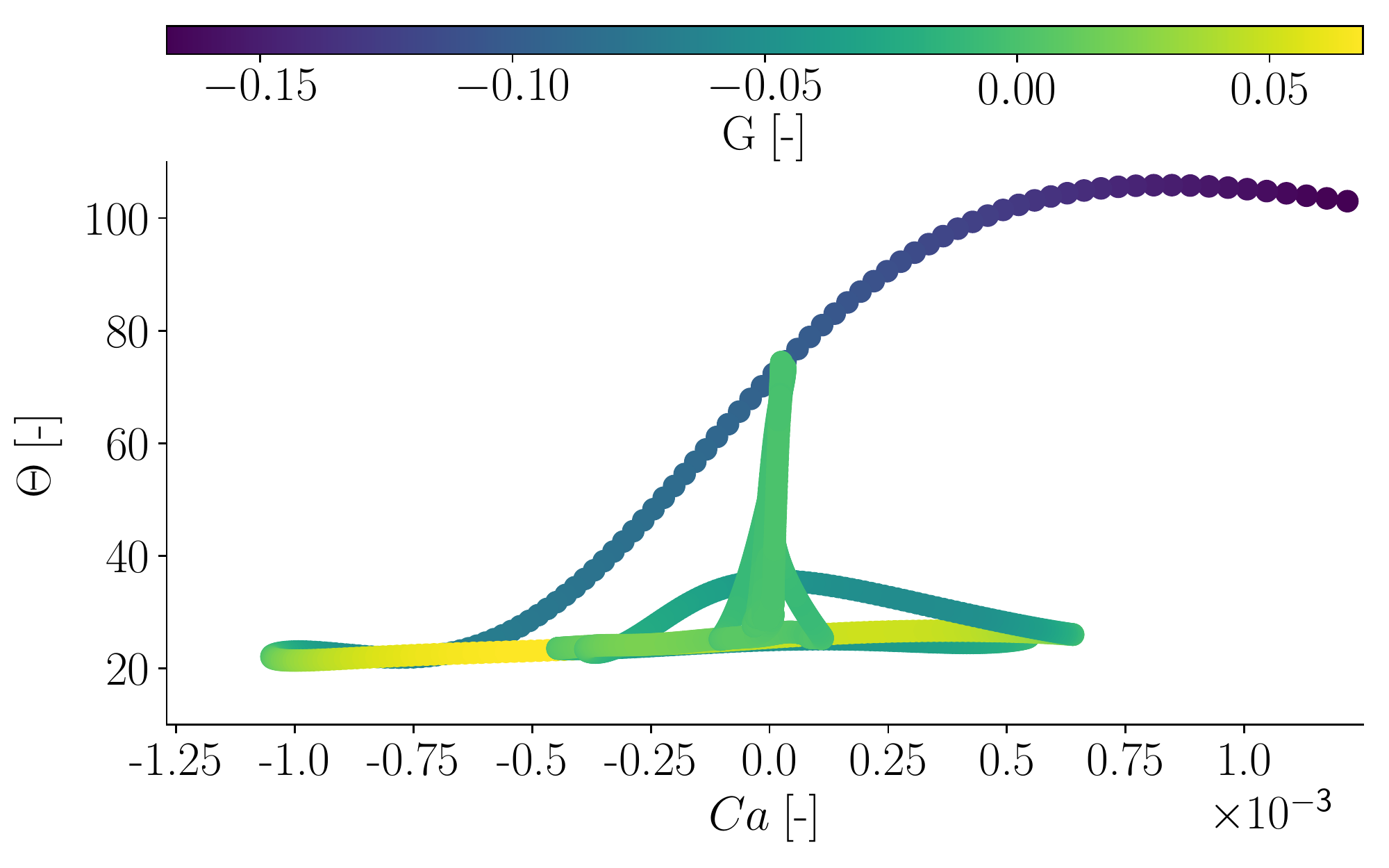}}

     \subfloat[]{\label{fig:model_prediction_comparison_1500}
       \includegraphics[width=.47\textwidth]{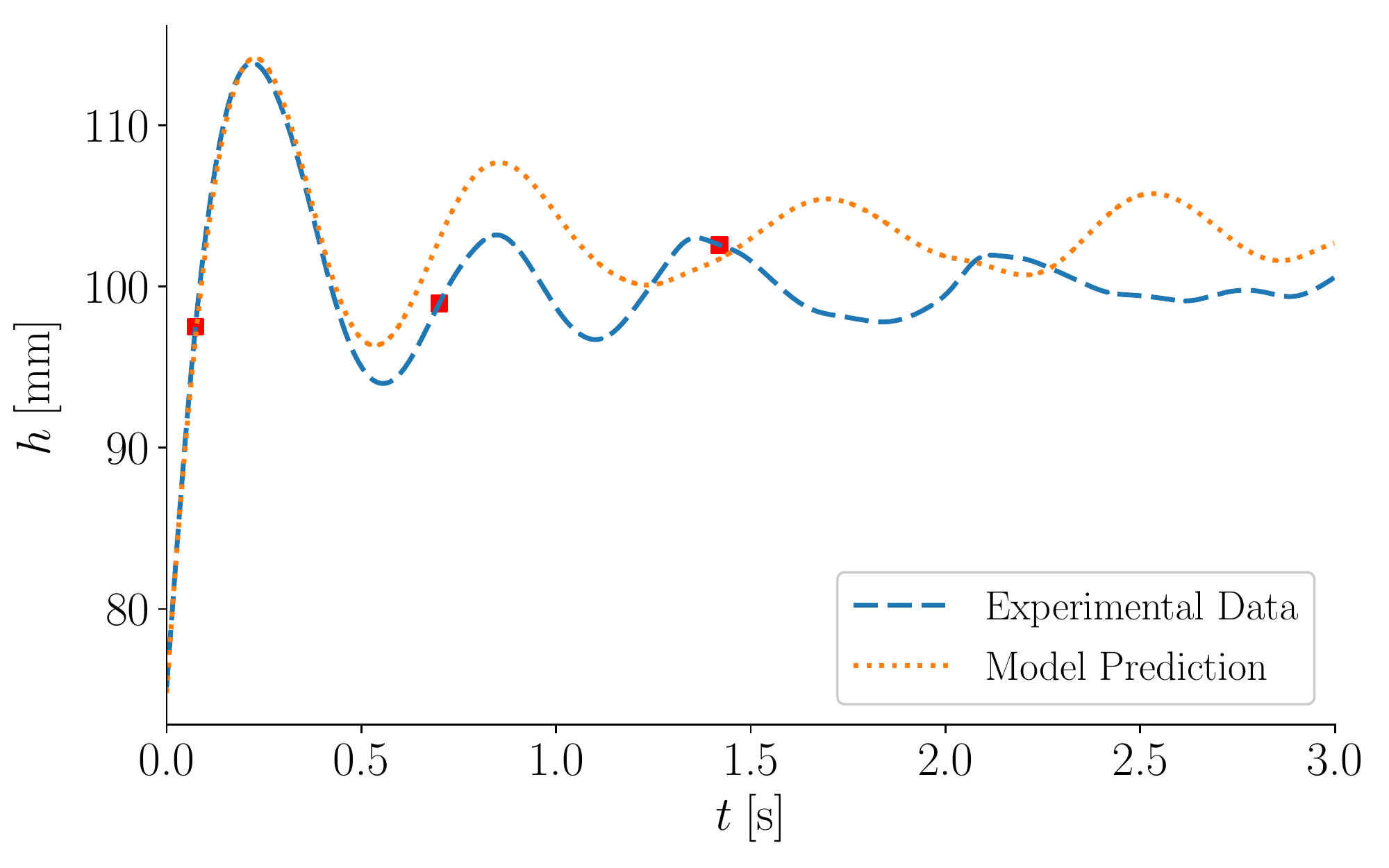}}
     \subfloat[]{\label{fig:theta_over_ca_1500}
       \includegraphics[width=.47\textwidth]{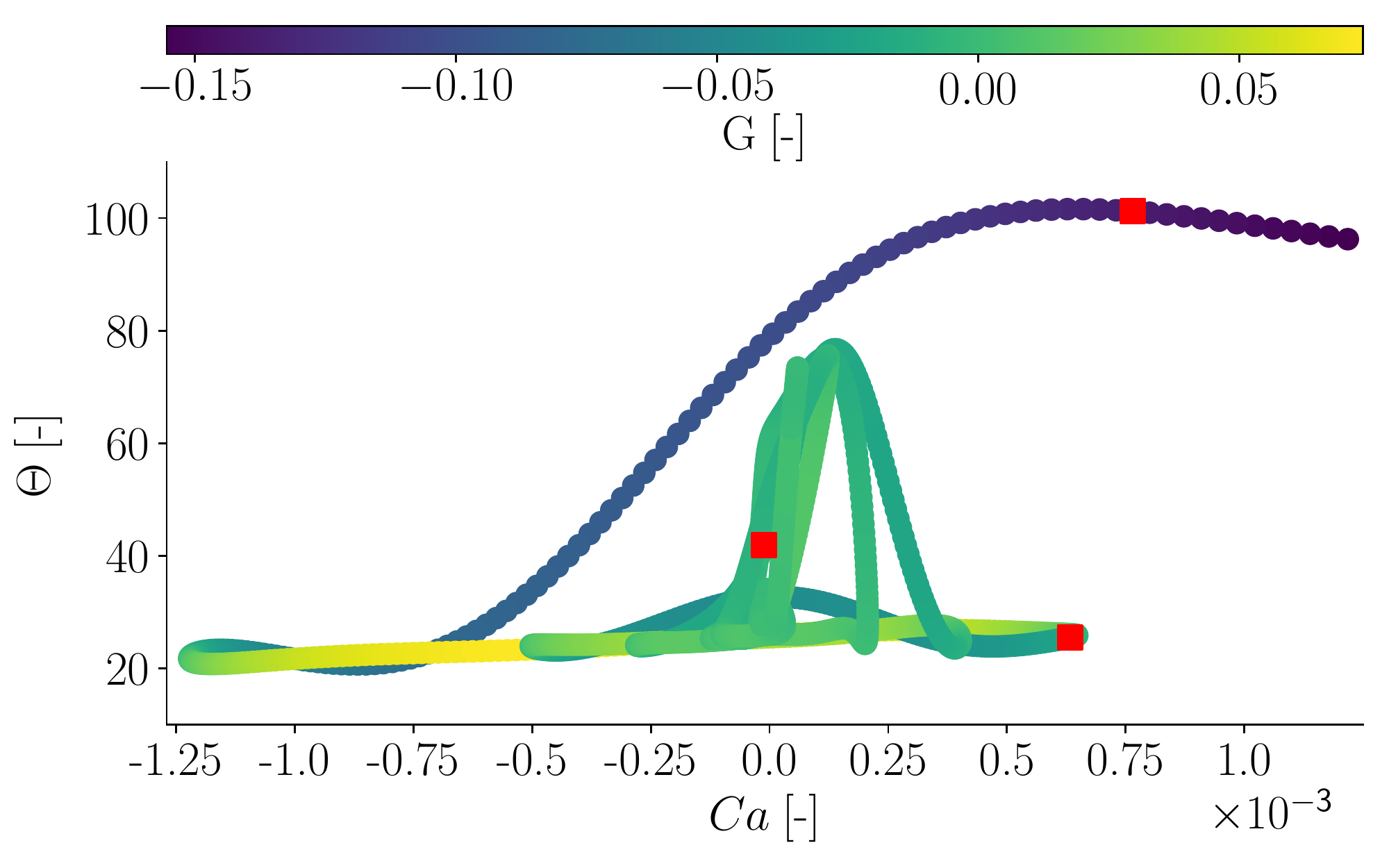}}

\caption{\textcolor{black}{LIF results of the oscillating interface for different test cases. Fig (a), (c) and (e) temporal evolution of the mean interface height inside the channel for Test Case 1, 2 and 3 respectively}. The dashed line represents the height taken from the images, the dotted line the predicted height from the model in Section \ref{sec:Model}. Fig (b), (d) and (f) show the measured contact angle as a function of capillary number $Ca=u_c\mu/\sigma$. The markers are colored according to the acceleration number $G$. The capillary number at $t = \SI{0}{\second}$ is $Ca = 5.3\times10^{-3}$, but the axis is limited in the range $[-1.25,1.25]\times 10^{-3}$ to better visualize the final region at $Ca\approx 0$. The red squares in Fig. (f) correspond to images in Figure \ref{fig:rise_interface} (c.f the times label in the captions of that figure).}\label{fig:ledar_results}
\end{figure*}

 We begin by illustrating the results of the interface detection via high speed LIF-based visualization. \textcolor{black}{The results of the repeated experiments yielded similar results, thus the results always show the case of the first repetition. Figure \ref{fig:ledar_results} shows three snapshots of Test Case 1, each taken at different points of the oscillation. Figure \ref{fig:falling_interface} shows three snapshots of Test Case 4.}
 In each figure, the text in the image recalls the dimensionless conditions characterizing each snapshot, namely the dimensionless velocity of the contact line in terms of capillary number $Ca=\mu u_c/\sigma$ and dimensionless acceleration $G=a/g$ as well as the time passed after the meniscus entered the FOV. \textcolor{black}{The signs of $Ca$ and $G$ follow the coordinate system defined in Fig. 2, i.e. a positive sign indicates an upwards motion or an acceleration upwards.} The results of the interface fitting, following equation \eqref{cosh} is shown using a dashed line.

 Despite the widely different conditions in terms of velocity and acceleration, the shape of the interface is well described by the model in equation \eqref{cosh}. The interface remains symmetric with respect to the vertical axis, while the contact angle in dynamic conditions largely varies between the rising and the descending conditions.

 \textcolor{black}{For Test Case 1 to 3, Figure \ref{fig:rise} plots on the left the evolution of the liquid height as a function of time (blue dashed lines) together with the prediction of the simple model in section \ref{sec:Model}}. These test cases differ in the amplitude of the pressure step introduced, thus on the overall level of acceleration. Despite the simplifications involved, the model correctly predicts the interface's position and velocity in the first $0.5$ s, i.e. at the end of the first cycle. Afterwards, as the interface oscillates around its equilibrium conditions, the model prediction loses accuracy.

 Among the various terms in equation \eqref{eq:momentum}, it was found that the inertial contribution (left-hand side term) was the dominant one, followed by the viscous and the surface tension terms. Given the good performances of the simple model in equation \eqref{cosh} in representing the shape of the interface and thus the accurate prediction of the capillary contribution, it is clear that the main limitation of the model arises from the modelling of the viscous term. Although this was not within the scope of this work, this result should be considered for future works on unsteady capillary or quasi capillary channels (e.g. see
 \cite{washburn1921,levine1976}).

 Finally, on the right side of Figure \ref{fig:rise}, maps for every test case the evolution of the contact angle $\Theta$ as a function of the capillary number $Ca$ is shown. The markers are coloured by the level of acceleration $G$ in each point. The range of contact angles spanned within one experiment is $\Theta\in[20,110^o]$. The largest values are produced during the first rise of the interface when the acceleration is also the largest ($G=-0.24$). The range of capillary number observed during the experiments is $Ca\in[-1.25,5.30] \times 10^{-3}$ and is approximately the same in all test cases. Large capillary numbers are not shown in the plots to focus on the regions with small capillary numbers. During the descending phase, the capillary number varies between $Ca=-1.25\times 10^{-3}$ to $Ca\approx 0$; yet the contact angle remains approximately constant and equal to $\Theta\approx 22^o$.
 On the contrary, during the final phase of the experiment, when the interface oscillation are vanishing and the contact line is almost at rest, the contact angle exhibits the most prominent variation (in the range $\Theta\in [20,80] ^o$. In this phase, also the acceleration is approximately constant. While these results suggest that the history of the contact line dynamic might play a role in predicting the contact angle, it also brings the question of whether \textcolor{black}{it is possible to use classic contact line models of the form $\Theta=f(Ca,G)$ for low viscous fluids and in the presence of acceleration.}

 \subsection{TR-PIV Experiment Results} \label{sec:piv_results}

 \begin{figure*}[ht]
     \centering
     \subfloat[]{\label{fig:acc_rise}{
     	\includegraphics[width=0.47\textwidth ]{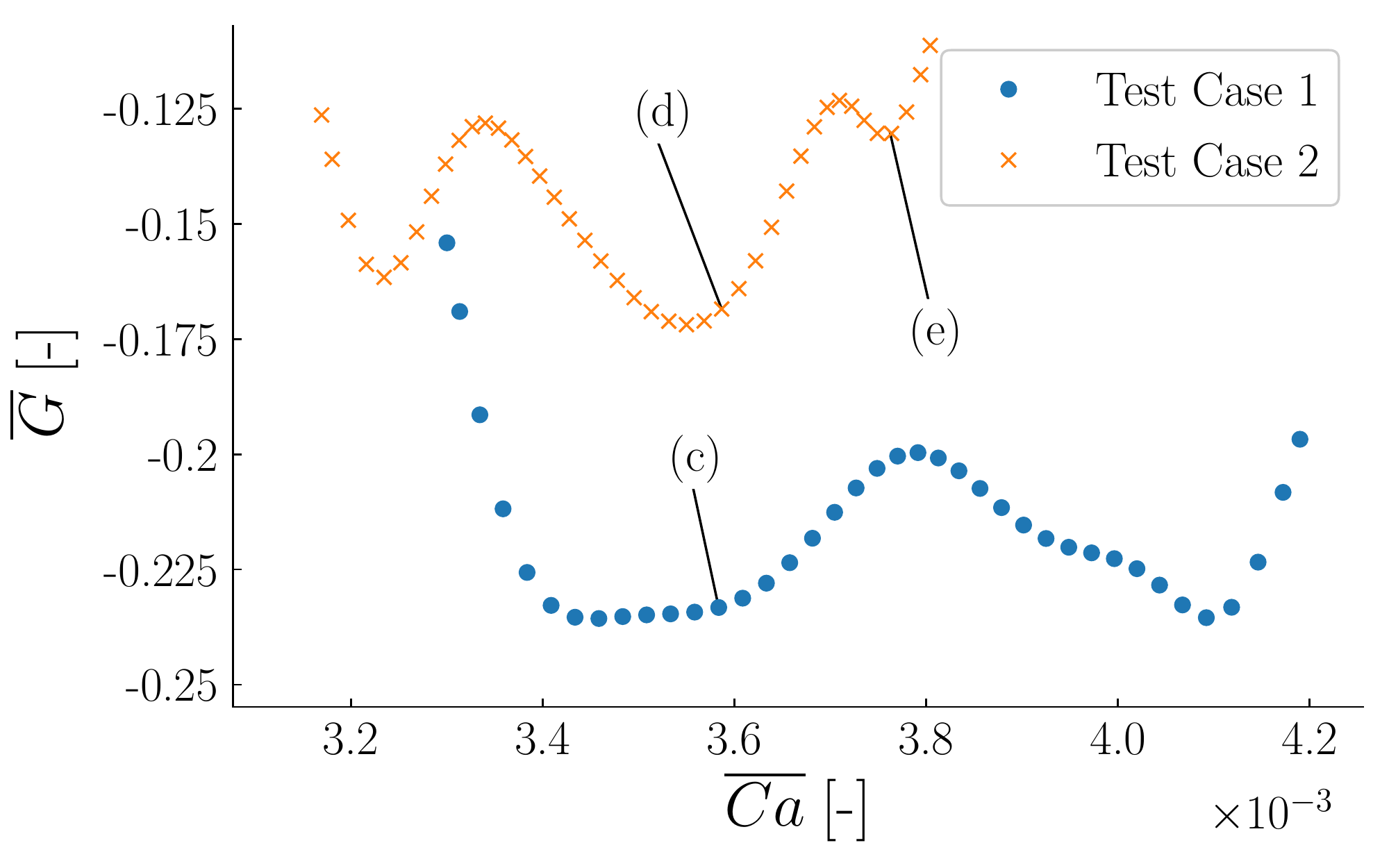}}
 	}
    ~
     \subfloat[]{\label{fig:qmax_rise}{
     	\includegraphics[width=0.47\textwidth ]{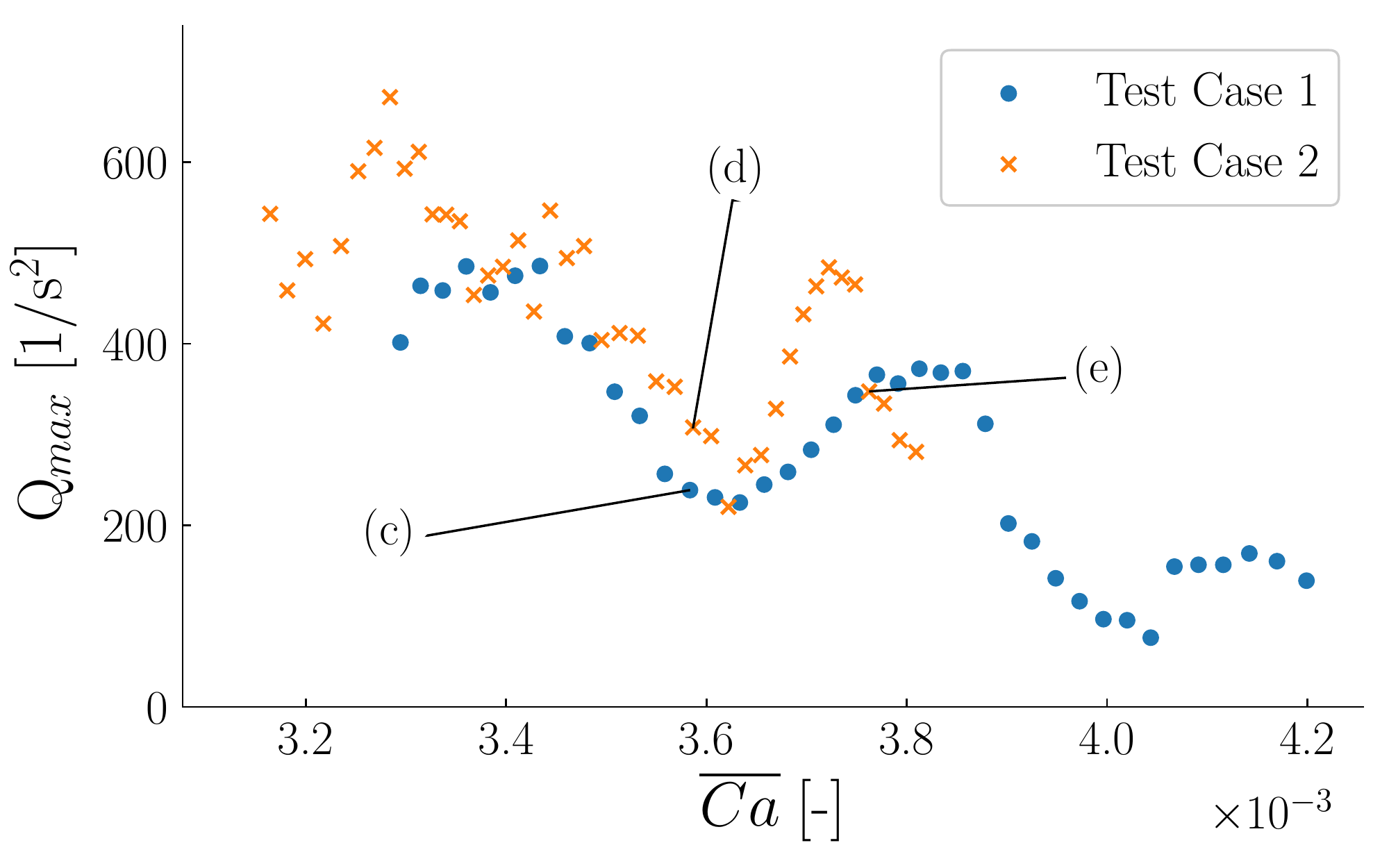}}
 	}
    
     \subfloat[]{\label{fig:rise_field_1}{
     	\includegraphics[width=.30\textwidth ]{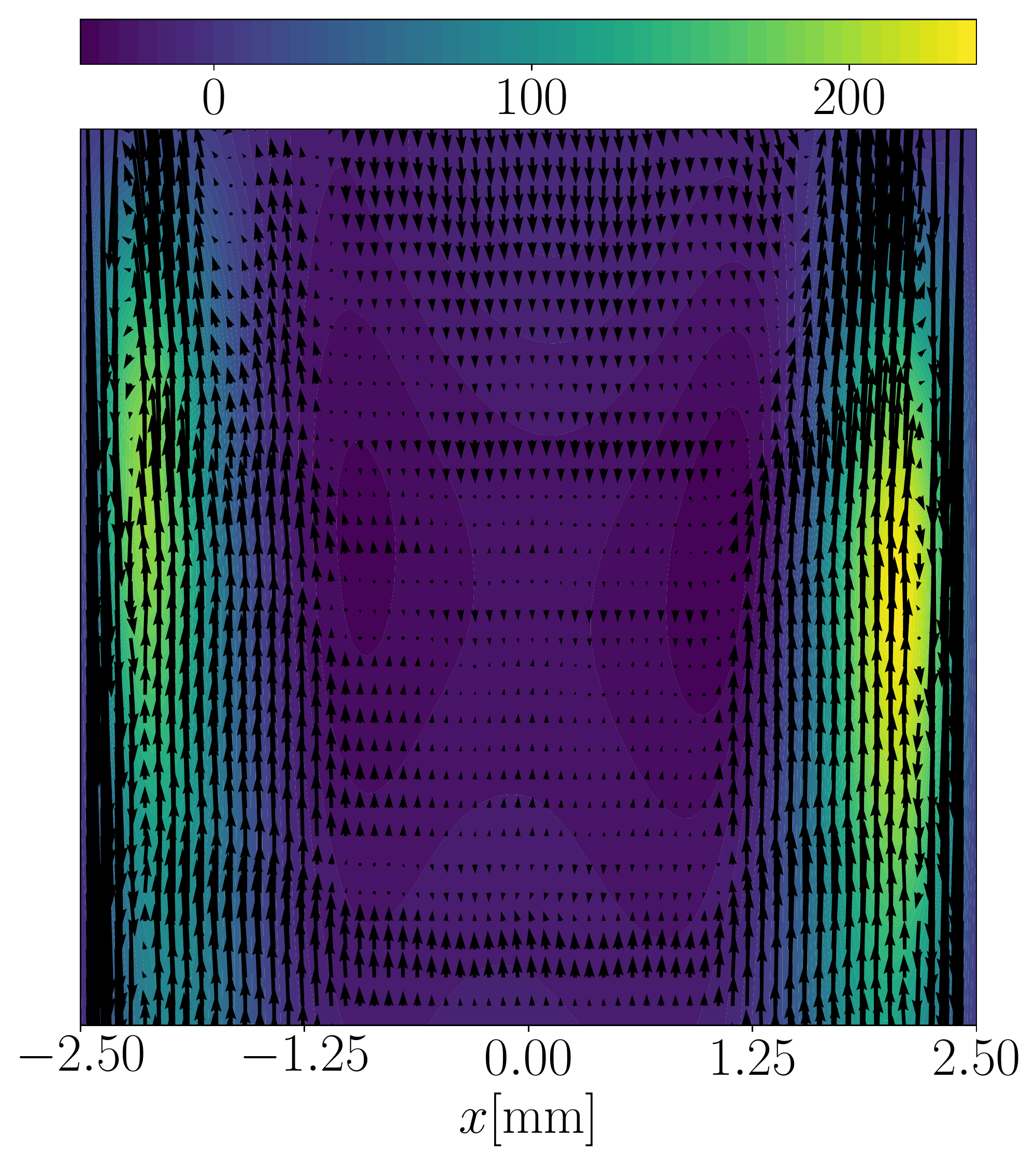}}
 	}
    ~
 	 \subfloat[]{\label{fig:rise_field_2}{
 		\includegraphics[width=.30\textwidth ]{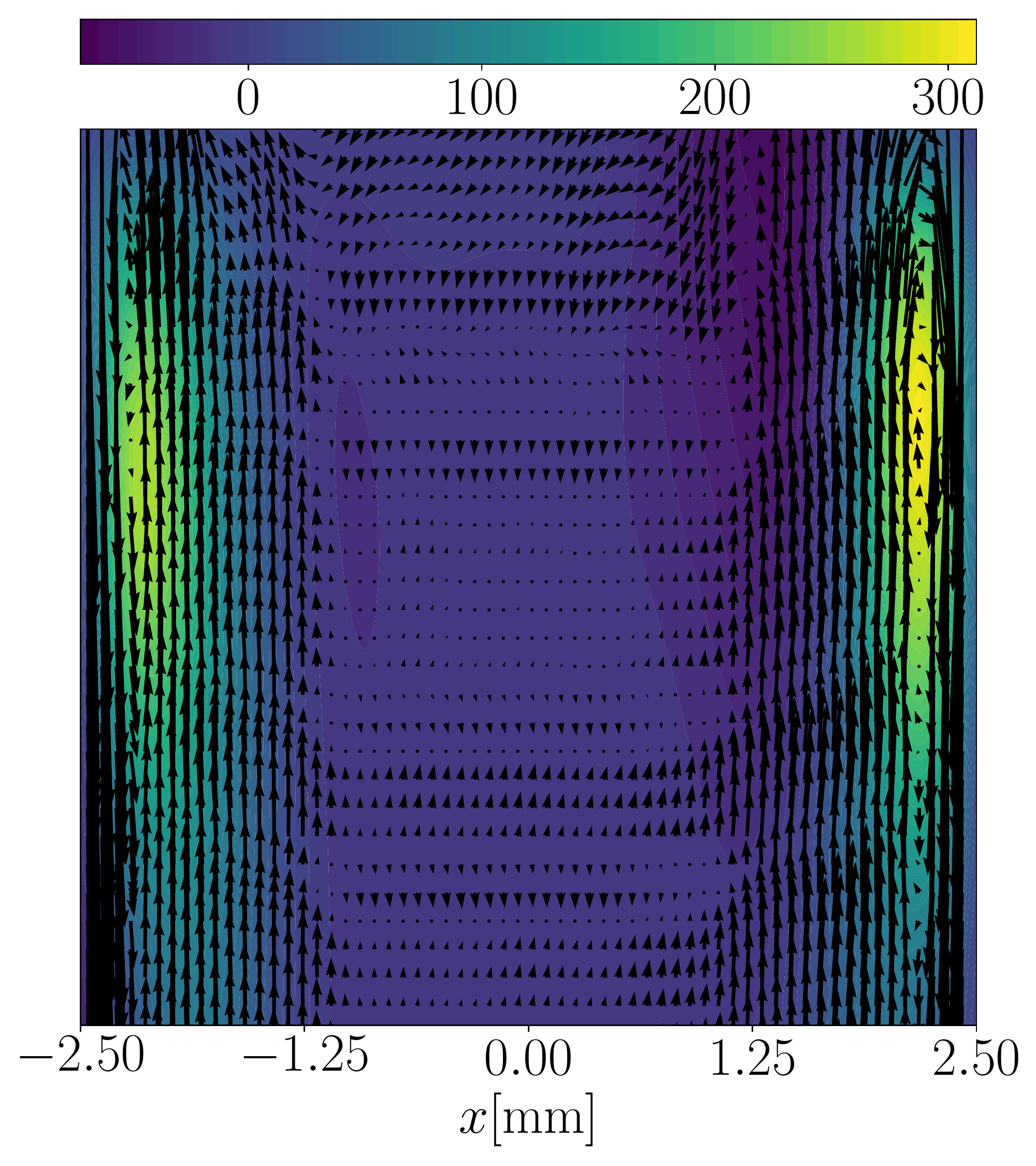}}
 	}
    ~
 	 \subfloat[]{\label{fig:rise_field_3}{
 	    \includegraphics[width=.30\textwidth ]{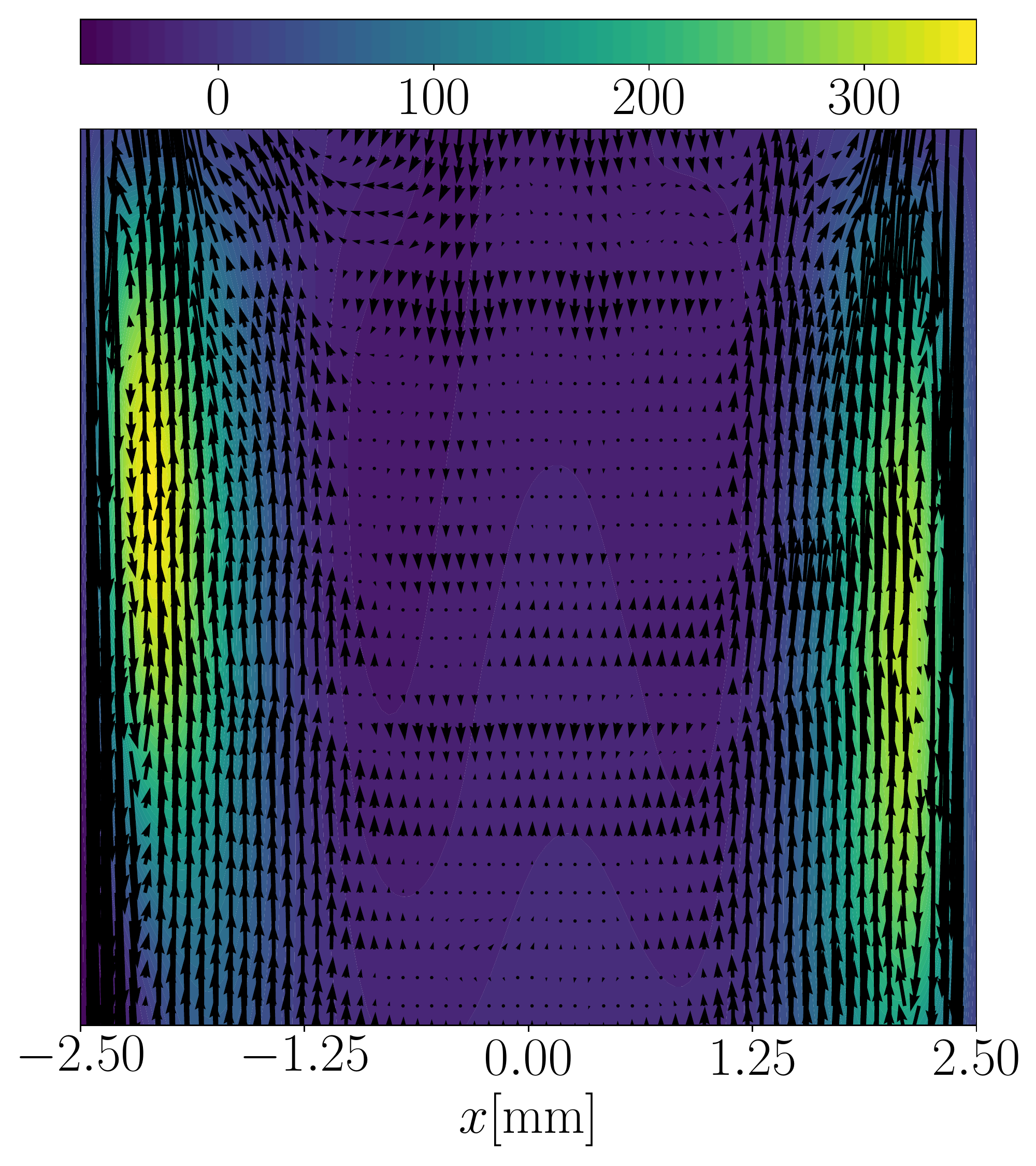}}
 	}
 	\caption{\textcolor{black}{PIV results for the oscillating interface}. Fig (a) - (b) dimensionless acceleration ($G$) and maximum $Q$-field as a function of the capillary number $Ca$ for two test cases. The blue circles correspond to Test Case 1 and the orange crosses to Test Case 2. Fig (c) - (e) $Q$-field contour and high-pass filtered velocity field. The aspect ratio of the plots is one.}
   \label{fig:rise}
 \end{figure*}

 \begin{figure*}[ht]
     \centering
 	\subfloat[]{\label{fig:acc_fall}{
     	\includegraphics[width=.47\textwidth ]{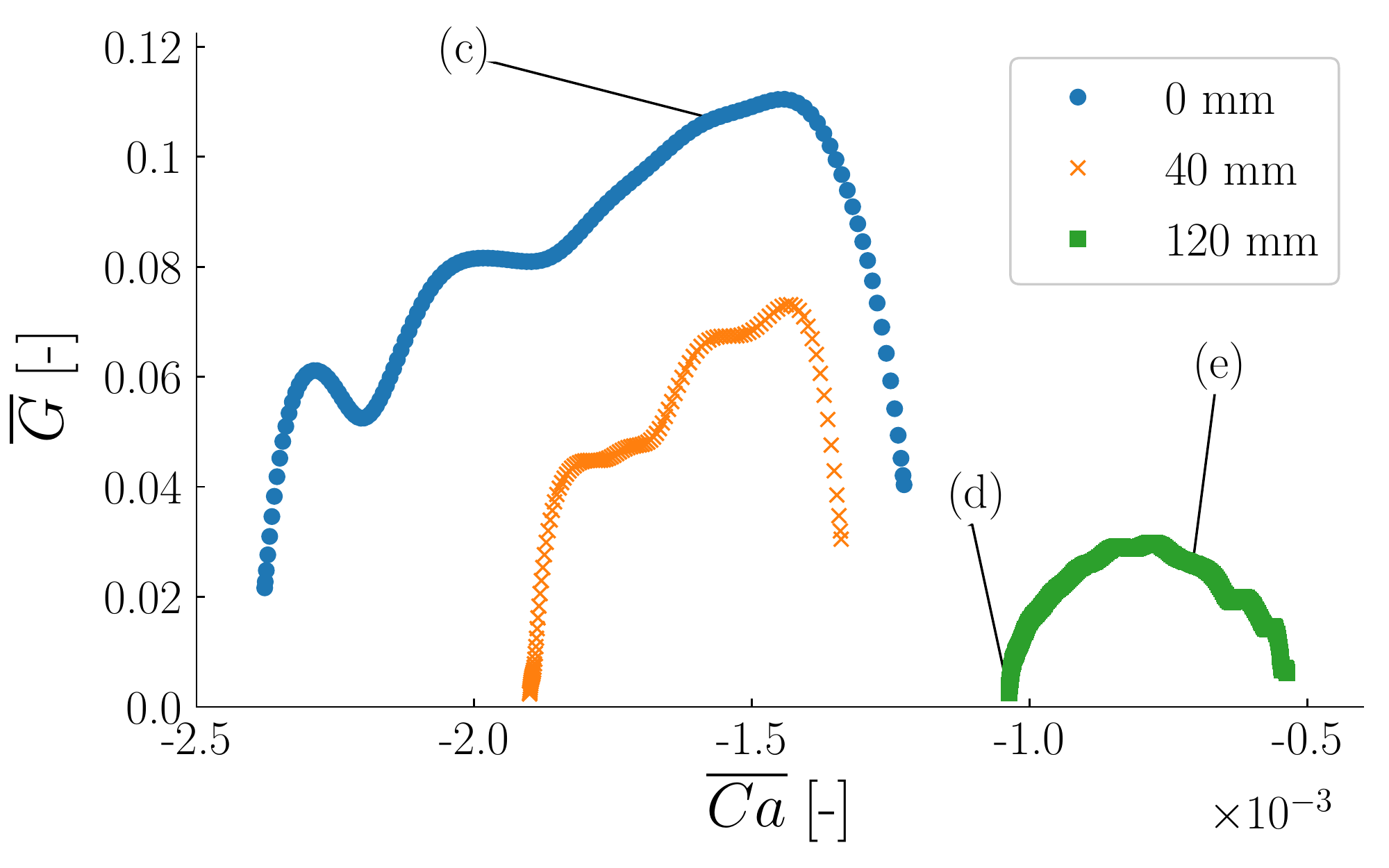}}
 	}
    ~
     \subfloat[]{\label{fig:qmax_fall}{
 	    \includegraphics[width=.47\textwidth ]{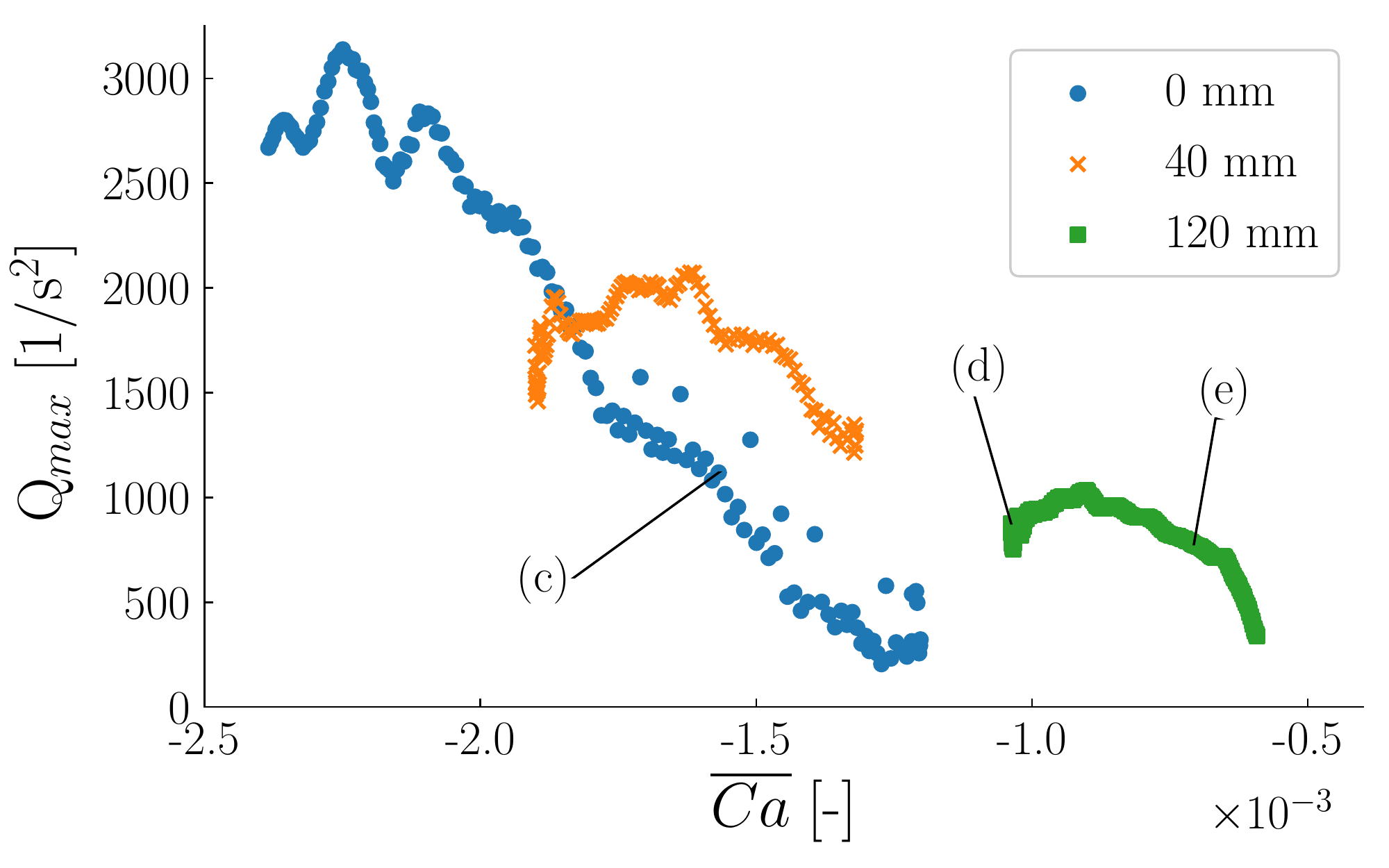}}
 	}
	
     \subfloat[]{\label{fig:fall_field_1}{
     	\includegraphics[width=.30\textwidth ]{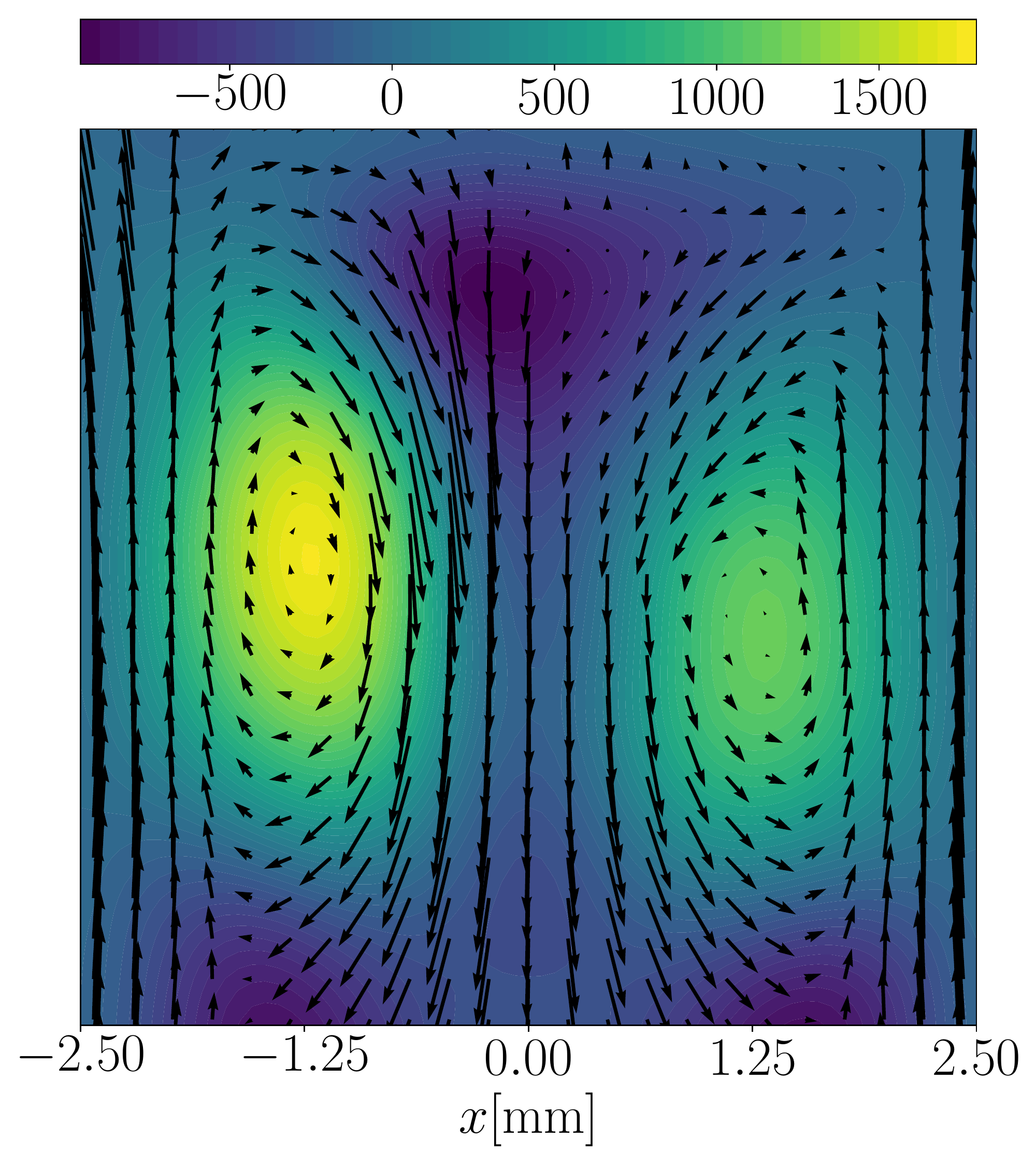}}
 	}
    ~
 	\subfloat[]{\label{fig:fall_field_2}{
 		\includegraphics[width=.30\textwidth ]{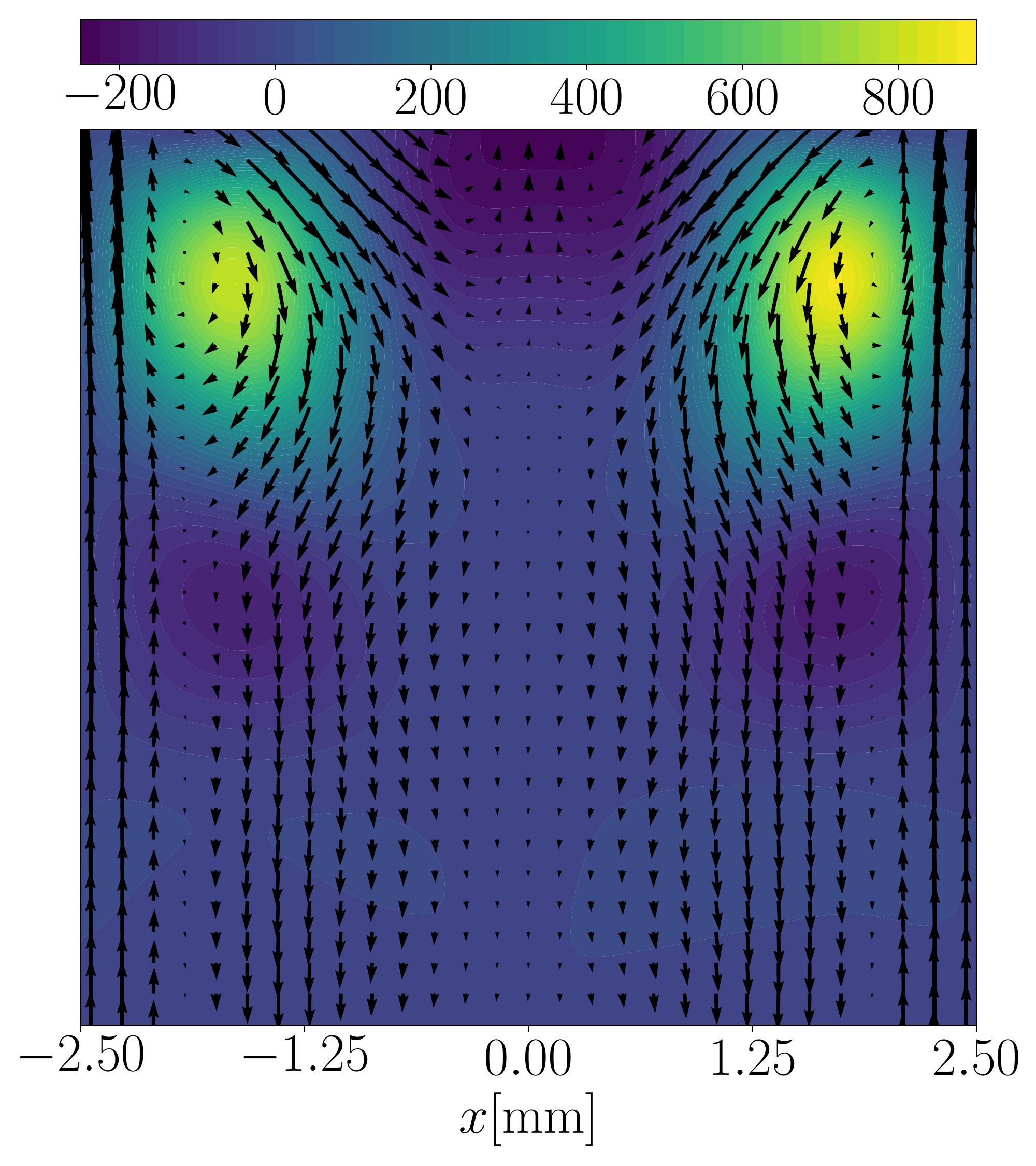}}
 	}
    ~
 	\subfloat[]{\label{fig:fall_field_3}{
 	    \includegraphics[width=.30\textwidth ]{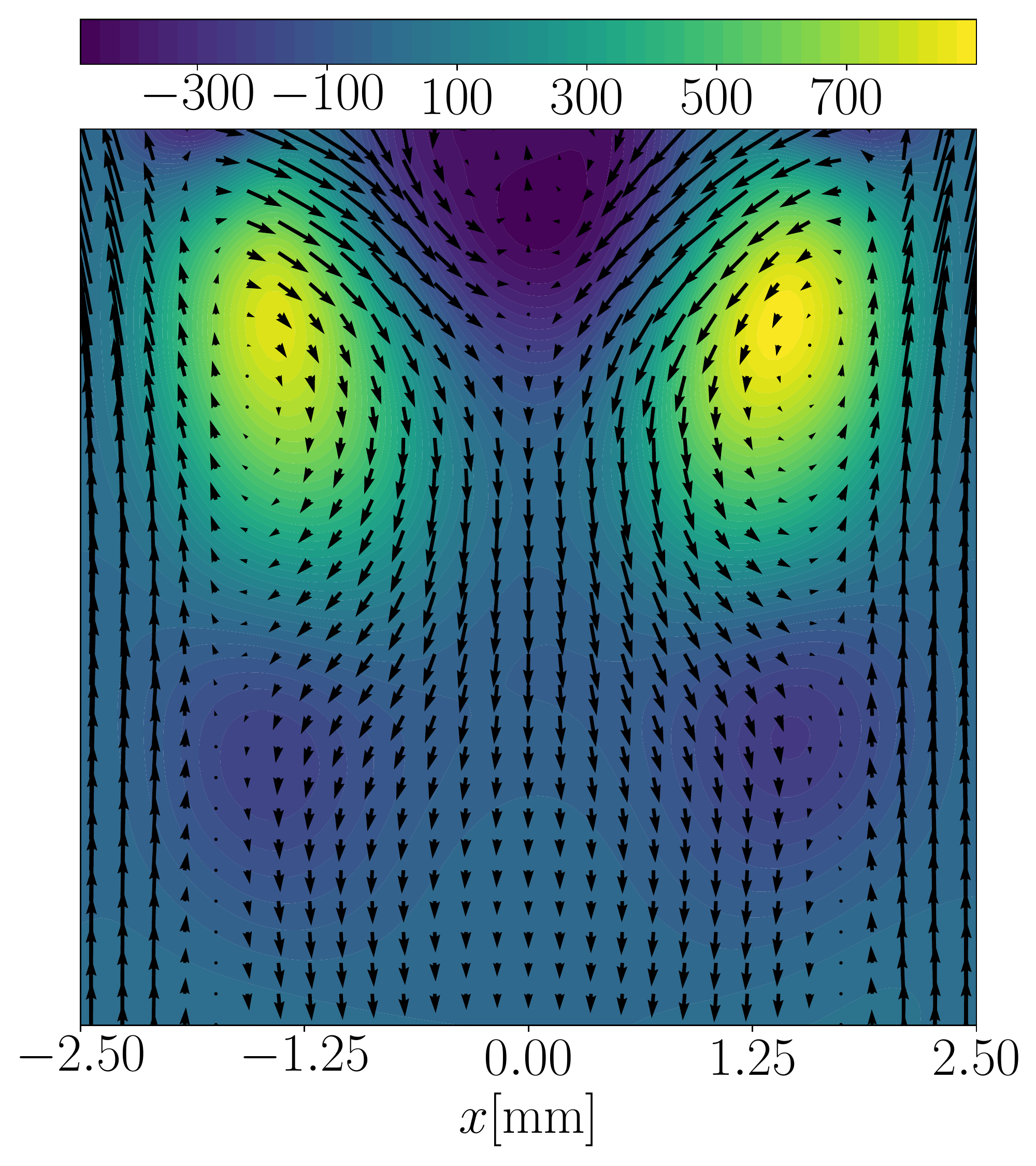}}
 	}
 	\caption{\textcolor{black}{Results for the descending interface. The figure is the same as Figure \ref{fig:rise}, but considering Test Case 3, 4 and 5}. The labels refer to the vertical position of the camera, where \SI{0}{\milli\meter} is the configuration in which the bottom boundary of the FOV is at the level of the reservoir's upper edge (see Figure \ref{fig:experimental_setup}). }
	\label{fig:fall}
 \end{figure*}

 The PIV velocity fields were processed retaining a total of $R=25$ POD modes in equation \eqref{POD_T}. The temporal structures of the modes were approximated with $n_\varphi=500$ RBFs with $\sigma_t = 0.01$ while $n_\phi=900$ RBFs were used for the spatial structures (see equation \eqref{eq:Gaussian_RBF}). For the regression in space, the RBFs were collocated over a regular grid of $n_{Bx} \times n_{By} = 60 \times 15$ and anisotropic kernels with $\sigma_x = 10$ and $\sigma_y = 15$ were selected. This allows to efficiently account for the high aspect ratio of the flow. No re-sampling was performed in the time domain since the available time resolution was sufficient for the scope of this work.

\textcolor{black}{Figure \ref{fig:rise} shows the results of Test Case 1 and 2, i.e. the advancing contact line}. Figure \ref{fig:acc_rise} and \ref{fig:qmax_rise} show the gravitational acceleration $G$ and the maximum of the $Q-$field representing the vortex strength within the ROI as a function of the capillary number $Ca$. The labels in the legend refer to the operating conditions described in Section \ref{sec:Model}. 
The capillary number in these plots \textcolor{black} {is different from the one used in the LIF campaign, i.e. it } is not based on the velocity of the contact line but on the mean velocity computed by zero-padding the velocity field at the walls and taking the mean of the average velocity of the bottom five rows. This operation acts as smoothing and serves for plotting purposes. \textcolor{black}{We distinguish these quantities with an over-bar in the labels of figure axes.}

A clear trend is visible: the $Q$-field increases with decreasing mean flow velocity.

 Three instantaneous velocity fields are shown in Figures \ref{fig:rise_field_1} - \ref{fig:rise_field_3}, corresponding to the three points marked in Figure \ref{fig:acc_rise} and \ref{fig:qmax_rise}. For plotting purposes, the velocity fields are slightly high-pass filtered by using a Gaussian filter with $\sigma_x = \sigma_y = 15$ and a truncation after $2.5$ standard deviations. The quiver plot only shows every third vector along $x$ and every second one along $y$ for better visibility. The $Q$-field is also shown in each snapshot for quantitative analysis, and the axis aspect ratio is set to one. We recall that the upper horizontal boundary of the ROI is at approximately \SI{1.5}{\milli\meter} from the interface at the centre of the channel. In each case, two counter-rotating vortices are visible close to the walls (the one on the right rotates clockwise and the one on the left counterclockwise). The vortices extend considerably along the vertical direction but only over a small distance in the cross-stream direction. Therefore, they have no remarkable influence on the velocity field at a distance of \SI{1}{\milli\meter} from the wall.

The initial pressure was high enough to produce a strong acceleration, as shown in Figure \ref{fig:acc_rise}. While the rolling motion of the flow is still present, the large acceleration and high velocities push the stream-line split injection pattern and its vortices towards the wall. It is worth noticing that the velocity fields look similar even though Figure \ref{fig:rise_field_1} and \ref{fig:rise_field_2} have largely different accelerations. Referring to the expected theoretical flow field from Figure \ref{fig:split_injection}, it appears that the stream-line split injection forms a smaller angle with the walls than what pictured in the schematic of Figure \ref{fig:split_injection}. As a result, the rolling motion expected by the viscous-capillary balance is confined to a narrow region of the flow.

\textcolor{black}{Figure \ref{fig:fall} shows the same results for Test Case 3 to 5}. The capillary and acceleration numbers in Figure \ref{fig:acc_fall} and \ref{fig:qmax_fall} have a different sign because the interface moves downward. A similar trend as for the rising interface can be observed, with the magnitude of the $Q$-field increasing as the velocity decreases. However, the reader should note that a decreasing velocity, in this case, means that the absolute value of the velocity is increasing. Another observation is the increased strength of the $Q$-field, with the maximum value being almost an order of magnitude above the maximum value for the rising interface. 

 The vortices in Figure \ref{fig:fall_field_1} - \ref{fig:fall_field_3} show a different behaviour than the ones for the rising interface. Besides the reversed rotation, as expected from the theoretical flow topology in Figure \ref{fig:split_stream}, their centres are much closer to the centre of the channel, and the rolling motion extends much further in both the $x$ and $y$ directions. In the case of the receding contact line, the stream-line split ejection (see Figure \ref{fig:split_ejection}) forms a larger angle with the wall than for the advancing contact line in the investigated configuration ($\Theta_0=33\pm2 ^o$).

The large difference cannot be justified by the different velocity, which in the slowest rising test cases is comparable to the fastest falling test cases, nor by its acceleration, which is also similar in the cases in \ref{fig:fall_field_1} and the one in Figure \ref{fig:rise_field_3}. Instead, the main difference between the rising and descending test cases is in the shape of the interface (cf. Figure \ref{fig:falling_interface} a) for a snapshot during the rise and Figure \ref{fig:falling_interface} c) for a snapshot during the fall). This impacts the velocity flow field in the entire ROI, thus more than \SI{5}{\milli\meter} below the interface.

 \section{Conclusions and Perspectives}
 \label{sec:conclusion}

 The dynamics of accelerating menisci was investigated using high-speed LIF-based visualization and Time-Resolved PIV.
 The first was used to track the evolution of the interface shape and the dynamic contact angle, while the second was used to measure the flow field in the fluid adjacent to the interface. Both advancing and receding contact lines were investigated for a wetting configuration (static contact angle $\Theta_0=33\pm2 ^o$) consisting of water flowing over acrylic glass.

 A simple model was used to describe the interface shape over the full range of operating conditions, with capillary numbers in the range $Ca\in[-1.25,5.30] \times 10^{-3}$ and contact lines acceleration in the range $G\in[-0.24,0.07]$. This model was also used to measure the dynamic contact angle which reaches values of $\Theta=110^o$ during a rise (at large accelerations) and $\Theta=20^o$ during a descent. Within the entire range of investigated conditions, it appeared impossible to relate the dynamic contact angle to the contact line velocity and/or acceleration: during some phases of the experiment, the contact angle remains constant over a wide range of $Ca$ and $G$, while in others it varies significantly over small intervals of $Ca$ and $G$. The quest for identifying a functional dependency $\theta=f(Ca,G)$ in inertia-dominated conditions might thus not be well posed.

 Concerning the velocity field underneath the interface, two counter-rotating vortices were observed below the menisci in both advancing and receding cases. \textcolor{black}{These were detected, and their intensity quantified in terms of $Q-$field, using a super-resolution approach combining RBFs and POD.} The topology of these vortices complies with the well known splitting streamline pattern postulated to solve the apparent incompatibility of contact line motion versus no-slip condition at the wall. The strength of these vortices depends on the mean velocity and the acceleration of the flow, and significantly differs between advancing and receding conditions. In the receding conditions, the splitting streamline forms a much larger angle with the wall and has a much larger impact on the flow field. Significant differences were observed between advancing/receding wetting configurations even when the modulus of velocity and acceleration of the contact lines were comparable. This highlights the relevance of capillary forces near the interface, since the main difference between the two cases was found in the interface's shape: this appeared nearly flat in the advancing cases and was characterized by large menisci during the receding test cases.

\bibliographystyle{spbasic}
\bibliography{Ratz_et_al_ECS_2022.bib}

\end{document}